\documentclass[submission,copyright]{eptcs}
\providecommand{\event}{GAM 2016} 	
\usepackage{breakurl}             	
\usepackage{graphicx}
\usepackage{wrapfig}
\usepackage{subcaption}
\usepackage{algorithmicx}
\usepackage{algpseudocode}
\usepackage{wrapfig}

\newcommand{\structure}[1]{}

\ifdefined \eg \else
	\newcommand{\eg}[1]{e.\,g.}
\fi

\ifdefined \ie \else
	\newcommand{\ie}[1]{i.\,e.}
\fi

\ifdefined \etal \else
	\newcommand{\etal}[0]{~et~al.~}
\fi

\ifdefined \cf \else
	\newcommand{\cf}[1]{cf.}
\fi

\ifdefined \footurl \else
	\newcommand{\footurl}[2]{\footnote{\url{#1} (last access: #2).}}
\fi

\ifdefined \refline \else
	\newcommand{\refline}[1]{l.~\ref{#1}}
\fi

\ifdefined \inlineparagraph \else
	\newcommand{\inlineparagraph}[1]{\noindent\textbf{#1}}
\fi

\ifdefined \figref \else
	\newcommand{\figref}[1]{Figure~#1}
\fi

\ifdefined \figabbrvref \else
\newcommand{\figabbrvref}[1]{Fig.~#1}
\fi

\ifdefined \secref \else
	\newcommand{\secref}[1]{Section~#1}
\fi

\ifdefined \secabbrvref \else
	\newcommand{\secabbrvref}[1]{Sec.~#1}
\fi

\ifdefined \chapref \else
	\newcommand{\chapref}[1]{Chapter~#1}
\fi

\ifdefined \chapabbrvref \else
	\newcommand{\chapabbrvref}[1]{Chap.~#1}
\fi

\ifdefined \appendixref \else
	\newcommand{\appendixref}[1]{Appendix~#1}
\fi

\ifdefined \tableref \else
	\newcommand{\tableref}[1]{Table~#1}
\fi

\ifdefined \equationref \else
	\newcommand{\equationref}[1]{Equation~#1}
\fi

\ifdefined \definitionref \else
	\newcommand{\definitionref}[1]{Definition~#1}
\fi

\ifdefined \algoref \else
	\newcommand{\algoref}[1]{Algorithm~#1}
\fi

\ifdefined \citepage \else
	\newcommand{\citepage}[2]{\mbox{\cite[p.\,#1]{#2}}}
\fi

\ifdefined \figemph \else
	\newcommand{\figemph}[1]{\textsf{#1}}
\fi

\ifdefined \stereotype \else
	\newcommand{\stereotype}[1]{\figemph{\guillemotleft #1\guillemotright}}
\fi

\ifdefined \codeindent \else
	\newcommand{\codeindent}{\hspace*{0.25cm}}
\fi

\newcommand{\hideHour}[1]{\IfStrEq{#1}{0}{}{\zfill{#1}{2}\,h~}}

\newcommand{\hideMin}[1]{\IfStrEq{#1}{0}{}{\zfill{#1}{2}\,min~}}

\newcommand{\hideSec}[1]{\IfStrEq{#1}{0}{}{\zfill{#1}{2}\,s}}

\newcommand{\zfill}[2]{%
	\StrLen{#1}[\mystringlen]%
	\ifnum \mystringlen < #2
	0#1%
	\else
	#1%
	\fi
}

\ifdefined \exectime \else
	\newcommand{\exectime}[3]{\hideHour{#1}\hideMin{#2}\hideSec{#3}}
\fi

\ifdefined \mem \else
	\newcommand{\mem}[1]{#1\,MB}
\fi
\usepackage{marginnote}
\usepackage[yyyymmdd,hhmmss]{datetime}

\ifdefined \structure \else
\newcommand{\structure}[1]{\marginpar{\tiny\textbf{#1}}}
\fi

\ifdefined \compiletime \else
	\newcommand{\compiletime}[0]{\today~\currenttime}
\fi
\usepackage{todonotes}

\ifdefined \tinytodo \else
	\newcommand{\tinytodo}[1]{\todo[backgroundcolor=white,linecolor=black,size=\tiny,noline]{#1}}
\fi

\ifdefined \tinytodolabel \else
\newcommand{\tinytodolabel}[2]{\label{#1}\todo[backgroundcolor=white,linecolor=black,size=\tiny,noline]{#2}}
\fi

\ifdefined \tinyinlinetodo \else
	\newcommand{\tinyinlinetodo}[1]{\todo[backgroundcolor=white,linecolor=black,size=\small,noline,inline]{#1}}
\fi

\ifdefined \tinyinlinetodolabel \else
\newcommand{\tinyinlinetodolabel}[2]{\label{#1}\todo[backgroundcolor=white,linecolor=black,size=\small,noline,inline]{#2}}
\fi

\algrenewcommand\alglinenumber[1]{\tiny #1:}
\algrenewcommand\algorithmicindent{.5em}

\title{Incremental View Maintenance for Deductive Graph Databases Using Generalized Discrimination Networks}
\author{Thomas Beyhl
\institute{Hasso Plattner Institute\\
at the University of Potsdam\\
Potsdam, Germany}
\email{thomas.beyhl@hpi.de}
\and
Holger Giese
\institute{Hasso Plattner Institute\\
at the University of Potsdam\\
Potsdam, Germany}
\email{holger.giese@hpi.de}
}
\def\titlerunning{Incremental View Maintenance for Deductive Graph Databases}
\def\authorrunning{T. Beyhl \& H. Giese}

\begin{document}
\maketitle

\begin{abstract}
Nowadays, graph databases are employed when relationships between entities are in the scope of database queries to avoid performance-critical join operations of relational databases.
Graph queries are used to query and modify graphs stored in graph databases.
Graph queries employ graph pattern matching that is NP-complete for subgraph isomorphism.
Graph database views can be employed that keep ready answers in terms of precalculated graph pattern matches for often stated and complex graph queries to increase query performance.
However, such graph database views must be kept consistent with the graphs stored in the graph database.

In this paper, we describe how to use incremental graph pattern matching as technique for maintaining graph database views.
We present an incremental maintenance algorithm for graph database views, which works for imperatively and declaratively specified graph queries.
The evaluation shows that our maintenance algorithm scales when the number of nodes and edges stored in the graph database increases.
Furthermore, our evaluation shows that our approach can outperform existing approaches for the incremental maintenance of graph query results.
\end{abstract}

\section{Introduction}
\structure{Review 1b}
Nowadays, graph databases are employed when relationships between entities are in the scope of graph database queries, because graph databases \textit{can} outperform relational databases, due to the fact that in graph databases a traversal from one node to an adjacent node is a constant time operation \cite{Rodriguez:2010aa}, in contrast to relational databases that require performance critical join-operations to traverse from one node to an adjacent node.
%
%
%
\structure{Review 1d}
%
%
%
%
Graph database queries employ graph pattern matching that is NP-complete for subgraph isomorphism \cite{Garey:1979aa}.
%
Thus, the graph query evaluation can be very time-consuming in worst-case scenarios.
One possibility to improve the performance of graph query evaluation is to employ graph database views that keep ready precalculated answers for graph queries.
%
\structure{Review 2a (3 sentences)}
These graph database views store all matches for graph patterns implemented by graph queries.
However, such graph database views must be kept consistent with the graphs stored in the  graph database.
When graphs in the graph database change, old matches that do no satisfy a certain graph pattern anymore must be removed from the graph database view and new matches that satisfy a certain graph pattern must be added to the graph database view.
Otherwise, graph queries lead to different graph query results when they make use of inconsistent graph database views.
%
\structure{Review 1a (1 sentence)}
Furthermore, Harrison\etal{}\cite{Harrison:1992aa} state that \textit{``in a deductive database, the task of maintaining materialized views is challenging, because views can be defined using negation and recursion''}.
We refer to graphs stored in graph databases as \textit{base graphs} and to graph database views derived from base graphs as \textit{view graphs}.
%

In this paper, we describe how to use incremental graph pattern matching as technique for maintaining view graphs.
The main contributions are a) an enumeration mechanism that enables to store graph pattern matches in view graphs in a manner that graph pattern matches can be reused effectively and can be maintained efficiently, b) an incremental maintenance algorithm for view graphs that works for imperatively and declaratively specified graph queries and scales when the number of nodes and edges in base graphs increases, and c) an evaluation which shows that our approach \textit{can} outperform existing approaches for incremental maintenance of view graphs.

\secabbrvref{\ref{sec:state_of_the_art}} describes the state of the art in incremental graph pattern matching.
\secabbrvref{\ref{sec:example}} introduces the running example used throughout this paper.
\secabbrvref{\ref{sec:views}} describes our concept of view graphs.
%
\secabbrvref{\ref{sec:algorithm}} describes how view graphs are created and maintained using incremental graph pattern matching.
%
\secabbrvref{\ref{sec:evaluation}} evaluates the performance of our approach.
\secabbrvref{\ref{sec:related_work}} compares related work with our approach.
\secabbrvref{\ref{sec:conclusion}} concludes our paper and outlines future work.

\advance\textheight 13.6pt
\advance\textheight 13.6pt

\section{State of the Art}
\label{sec:state_of_the_art} 
In practice, discrimination networks are used to maintain large collections of working memory elements that satisfy certain conditions.
For example, discrimination networks are widely used in active database management systems \cite{Hanson:1996aa}, view maintenance for relational databases \cite{Hanson:2002aa} and incremental graph pattern matching \cite{Bergmann:2008jk}.
Several kinds of discrimination networks exist such as Rete networks \cite{Forgy:1982cg}, TREAT \cite{Miranker:1987tr}, and Gator networks \cite{Hanson:2002aa}.
All kinds of discrimination networks consist of network nodes and edges that constitute a directed acyclic network structure, but differ in the kinds of employed network nodes.
%
%
In general, a network node performs a condition test to check whether working memory elements (e.\,g., tuples of relational data or nodes of graph data) satisfy a certain condition and, afterwards, store which working memory elements satisfy this condition.
Network edges describe the exchange of working memory elements, which satisfy or dissatisfy conditions, between network nodes.
For graph pattern matching, network nodes employ graph conditions \cite{Ehrig:2004aa} as condition tests and store graph pattern matches that satisfy these graph conditions.
Furthermore, successor network nodes reuse graph pattern matches stored by predecessor network nodes for condition testing.
Therefore, each network node enumerates graph pattern matches that satisfy certain conditions.
From database perspective, these enumerations are considered as database views that keep ready graph pattern matches.
When the graph data changes, the changes are propagated through the network to update graph pattern matches stored by network nodes by re-evaluating graph conditions only for changed, added, and deleted nodes and edges of base graphs.

The most generalized kind of discrimination network is the Gator network \cite{Hanson:2002aa} that allows network nodes with an arbitrary number of inputs.
Rete networks \cite{Forgy:1982cg} and TREAT networks \cite{Miranker:1987tr} are extreme examples of Gator networks due to the following restrictions.
Rete networks are limited to network nodes with at most two inputs.
Furthermore, Rete networks must be either left- or right-associative \cite{Lee:1992rt} and must not consist of re-convergent network nodes, otherwise the original Rete match algorithm produces duplicated or missing matches \cite{Lee:1992rt}.
TREAT networks are restricted to network nodes with at most one input and do not allow intermediate nodes.
In TREAT, join conditions are computed on demand.
%

Bunke\etal \cite{Bunke:1991cw} transferred the concepts of original Rete networks to the efficient implementation of graph grammars by deriving the Rete network from the left-hand sides of graph grammar rules.
Furthermore, EMF-IncQuery \cite{Bergmann:2008jk} employs Rete networks for incremental graph pattern matching in several application domains such as model queries over EMF models \cite{Bergmann:2010th},  derivation of features in EMF models \cite{Rath:2012aa}, live model transformations \cite{Rath:2008ex}, and synchronization of view models \cite{Debreceni:2014aa}.
However, current incremental graph pattern matching approaches are limited to Rete networks and, thus, do not allow arbitrary network structures although optimized generalized network structures such as Gator networks \cite{Hanson:2002aa} can outperform Rete networks in time and space as described by Hanson\etal \cite{Hanson:2002aa} for relational databases.
To our best knowledge, no approach exists that employs Gator networks as most generalized kind of discrimination network for incremental graph pattern matching and view maintenance of graph databases.
Discrimination networks do not support recursion due to their acyclic network structure.

\structure{Review 1e}
In general, Rete and Gator networks have the same expressiveness for graph pattern matching.
A formal proof is left for future work.
In contrast to current approaches for incremental graph pattern matching, our approach employs Gator networks as generalized kind of discrimination network to enable graph databases users to steer the tradeoff between memory consumption and time required to update the state of the discrimination network.
Computing optimal network structures is left for future work.
%
%

\section{Running Example}
\label{sec:example}
\begin{figure}[htbp]
	\centering
	\begin{subfigure}[b]{.3\linewidth}
		\centering
		\includegraphics[scale=0.5]{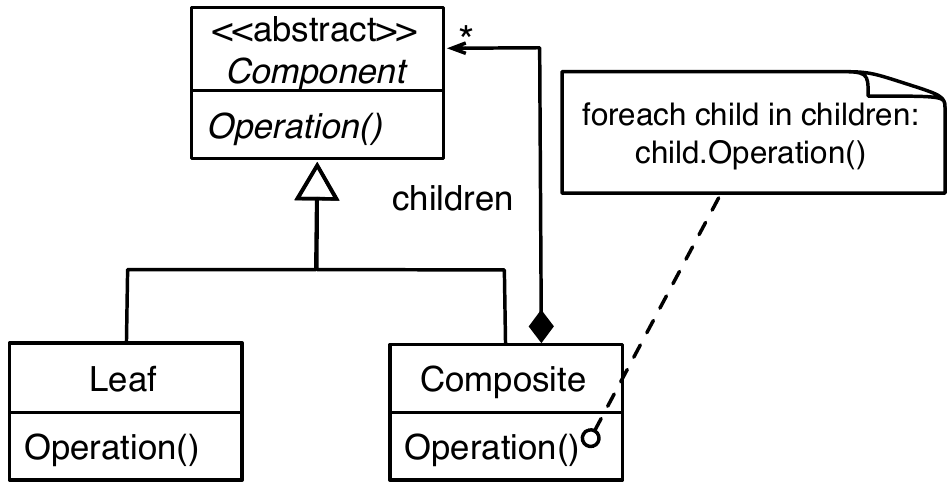}
		\caption{Composite design pattern \cite{Gamma:1994wx}}
		\label{fig:running_example_composite}
	\end{subfigure}
	\begin{subfigure}[b]{.44\linewidth}
		\centering
		\includegraphics[scale=0.5]{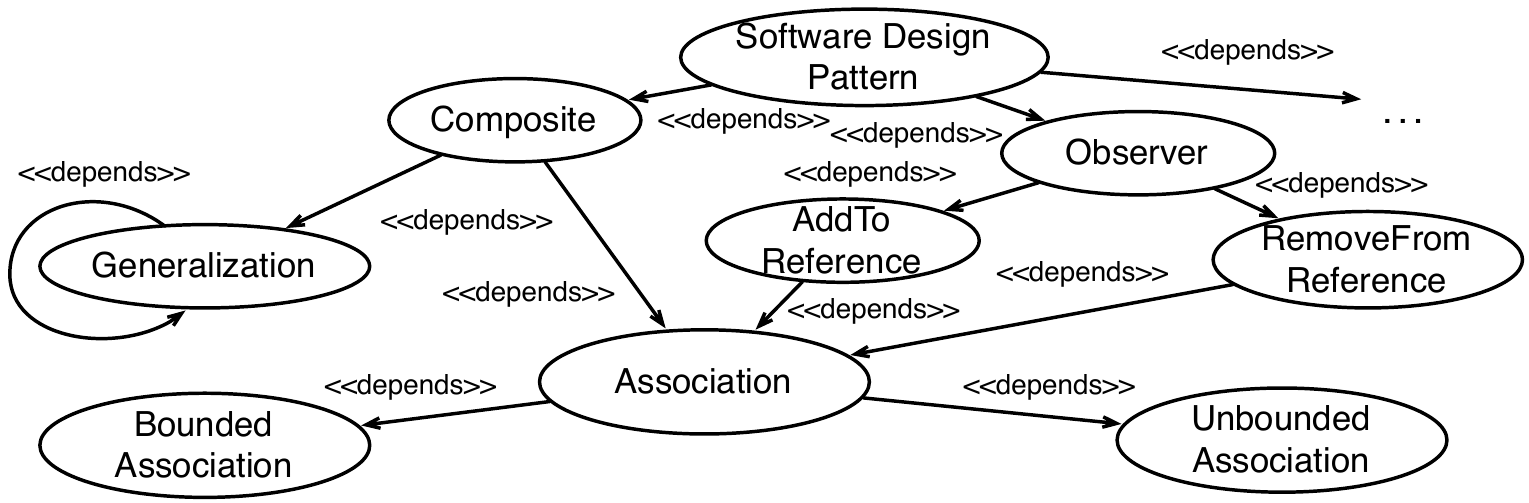}
		\caption{Dependency graph of graph properties}
		\label{fig:running_example_dependency_graph}
	\end{subfigure}
	\begin{subfigure}[b]{0.24\linewidth}
		\centering
		\includegraphics[scale=0.6]{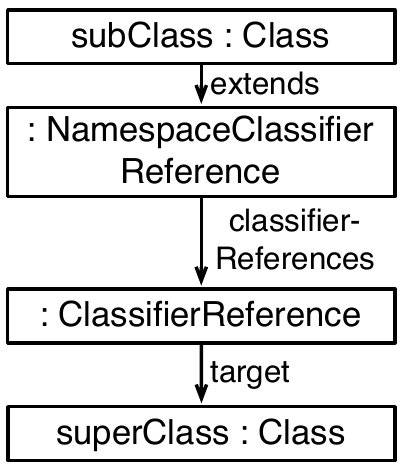}
		\caption{Generalization}
		\label{fig:running_example_pattern_generalization}
	\end{subfigure}
	\begin{subfigure}[b]{0.22\linewidth}
		\centering
		\includegraphics[scale=0.6]{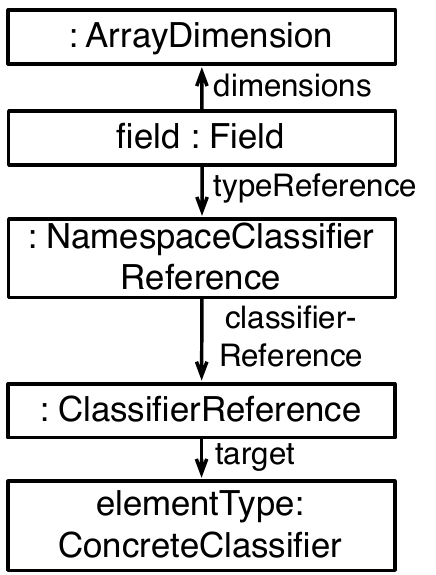}
		\caption{BoundedAssociation}
		\label{fig:running_example_pattern_arrayfield}
	\end{subfigure}
	\begin{subfigure}[b]{0.38\linewidth}
		\centering
		\includegraphics[scale=0.6]{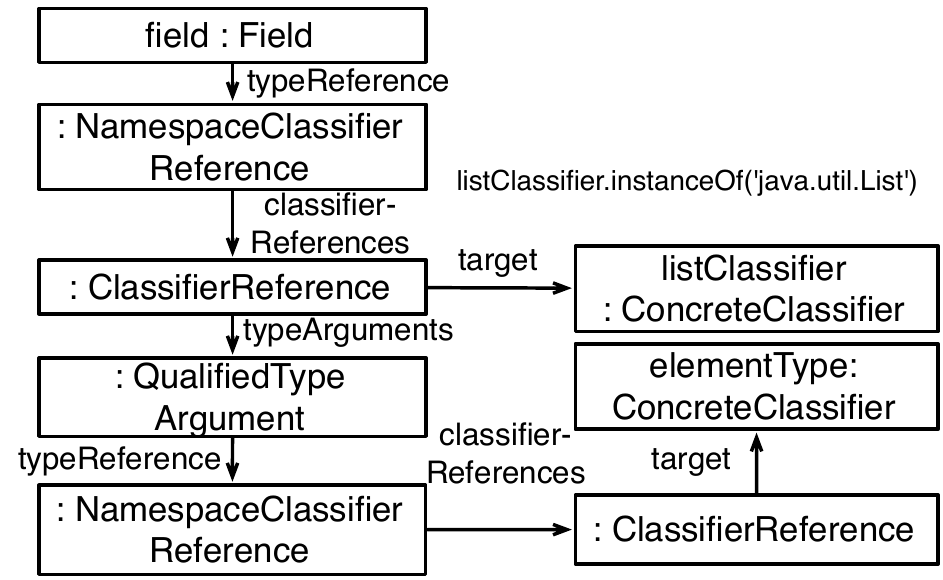}
		\caption{UnboundedAssociation}
		\label{fig:running_example_pattern_listfield}
	\end{subfigure}
	\begin{subfigure}[b]{0.28\linewidth}
		\centering
		\includegraphics[scale=0.6]{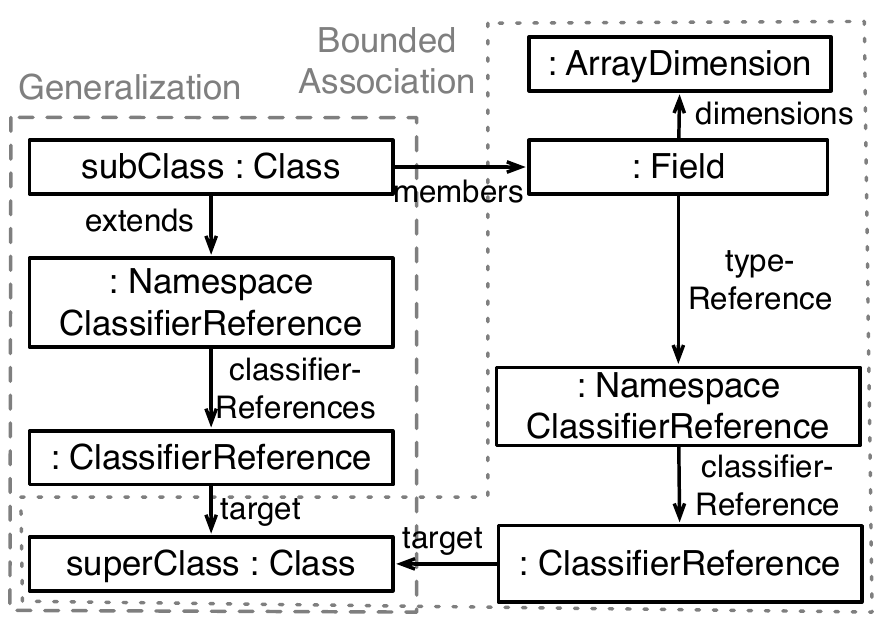}
		\caption{Composite}
		\label{fig:running_example_pattern_composite}
	\end{subfigure}
	\vspace{-4mm}
	\caption{Overview of running example with graph patterns} 
	\vspace{-4mm}
	\label{fig:running_example}
\end{figure}

%
%
%
%
\structure{Review 3c}
In our running example, we search for graph pattern matches that describe employed software design patterns \cite{Gamma:1994wx} in abstract syntax graphs (ASGs) of source code to analyze the evolution of software architectures.
We aim for an incremental maintenance of these graph pattern matches for employed software design patterns when ASGs change.
%
\structure{Review 3g}
High-level graph properties such as the Composite design pattern can be recovered by detecting low-level graph properties such as generalizations and associations \cite{Niere:2003uq}.
We employ graph pattern matching to detect such graph properties in terms of graph pattern matches in ASGs.
\figabbrvref{\ref{fig:running_example_composite}} shows the Composite design pattern as UML class model.
%
%
%
First, the \figemph{Composite} class must be a specialization of the \figemph{Component} class.
Second, the \figemph{Composite} class must own an association with the \figemph{Component} class as target.
\figabbrvref{\ref{fig:running_example_dependency_graph}} shows in a dependency graph that \figemph{Generalization}s and \figemph{Association}s have to be detected to recover \figemph{Composite} design patterns.
%
%
\figemph{Generalization}s depend on \figemph{Generalization}s, because multiple generalizations can constitute a multi-level generalization.
We distinguish \figemph{Association}s into \figemph{BoundedAssociation}s and \figemph{UnboundedAssociation}s, which employ data structures of immutable and mutable length to implement associations, respectively.
%

\structure{Review 3d}
\figabbrvref{\ref{fig:running_example}} shows the graph patterns used to detect \figemph{Generalization}s, \figemph{BoundedAssociation}s, \figemph{UnboundedAssociation}s, and \figemph{Composite} design patterns.
The \figemph{Generalization} graph pattern (\cf{} \figabbrvref{\ref{fig:running_example_pattern_generalization}}) describes that a sub\-class points via a namespace and classifier reference to a super\-class.
The \figemph{BoundedAssociation} graph pattern (\cf{} \figabbrvref{\ref{fig:running_example_pattern_arrayfield}}) describes that a field consists of an array dimension and points via a namespace and classifier reference to a classifier of elements stored in an array data structure.
The \figemph{UnboundedAssociation} graph pattern (\cf{} \figabbrvref{\ref{fig:running_example_pattern_listfield}}) describes that a field points via a namespace and classifier reference to a classifier that is an instance of a list data structure and, additionally, points via a qualified type argument, namespace and classifier reference to a classifier of elements stored in a list data structure.
The \figemph{Composite} graph pattern (\cf{} \figabbrvref{\ref{fig:running_example_pattern_composite}}) describes that a \figemph{Generalization} between a sub\-class and super\-class exists (\cf{} dashed polygon) and that a field of a sub\-class is a \figemph{BoundedAssociation} that points via a namespace and classifier reference to a super\-class of the generalization that describes the kind of the elements owned by the association (\cf{} dotted polygon).
%
Note, the \figemph{BoundedAssociation} graph pattern can be replaced by the \figemph{UnboundedAssociation} graph pattern.

\vspace{-2mm}
\section{Views for Deductive Graph Databases}
\label{sec:views}
\vspace{-2mm}
\structure{Basic problem definition (Review 2d)}
View graphs must store matches for graph patterns that are defined by graph database users.
Thus, the question arises what exactly are view graphs and how to specify their content.
View graphs must store graph pattern matches in a manner that a) graph pattern matches are stored memory-efficient and b) nodes with certain roles in graph pattern matches can be accessed effectively.
\secabbrvref{\ref{sec:view_graph}} describes our notion of view graphs.
\secabbrvref{\ref{sec:view_definition}} describes how our approach enables graph database users to define the content of view graphs.
%
%
We present an exhaustive description in our technical report \cite{Beyhl:2015tr}.


\subsection{Introduction of View Graphs}
\label{sec:view_graph}
Base graphs and view graphs are typed graphs that must be conform to a type graph.
Base graphs are graphs stored by graph databases to represent domain knowledge.
View graphs are graphs that store typed nodes and edges which together mark matches of graph patterns.
The node types in view graphs describe the kinds of graph pattern matches that are marked.
The edge types in view graphs describe the roles of nodes in marked graph pattern matches.
%
%
The edge types enable graph database users to effectively access nodes of graph pattern matches without the need to match them again to determine their role in graph pattern matches.
%
%
Furthermore, each node that participates in a graph pattern match is either referenced by an edge or scope.
When base graphs change in a way that nodes of view graphs do not mark valid graph pattern matches anymore, edges and scopes enable an efficient look-up of nodes in view graphs that must be revised.
This look-up is performed by traversing edges and scopes in backward direction from changed nodes of base graphs to nodes of view graphs that own these edges and scopes.
%

\begin{figure}[tbp]
	\centering
	\begin{subfigure}[b]{.49\linewidth}
		\centering
		\includegraphics[scale=0.51]{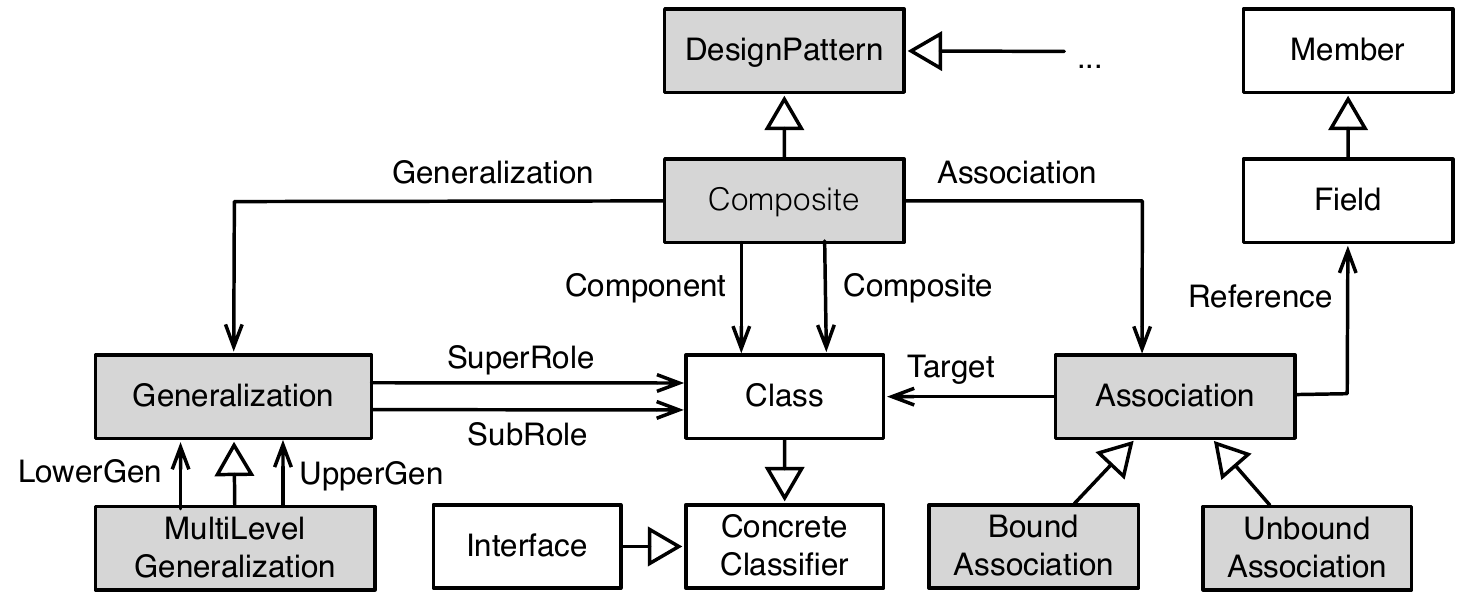}
		\caption{Base and view type graph as UML class model}
		\label{fig:view_reference_graph}
	\end{subfigure}
	\begin{subfigure}[b]{.49\linewidth}
		\centering
		\includegraphics[scale=0.55]{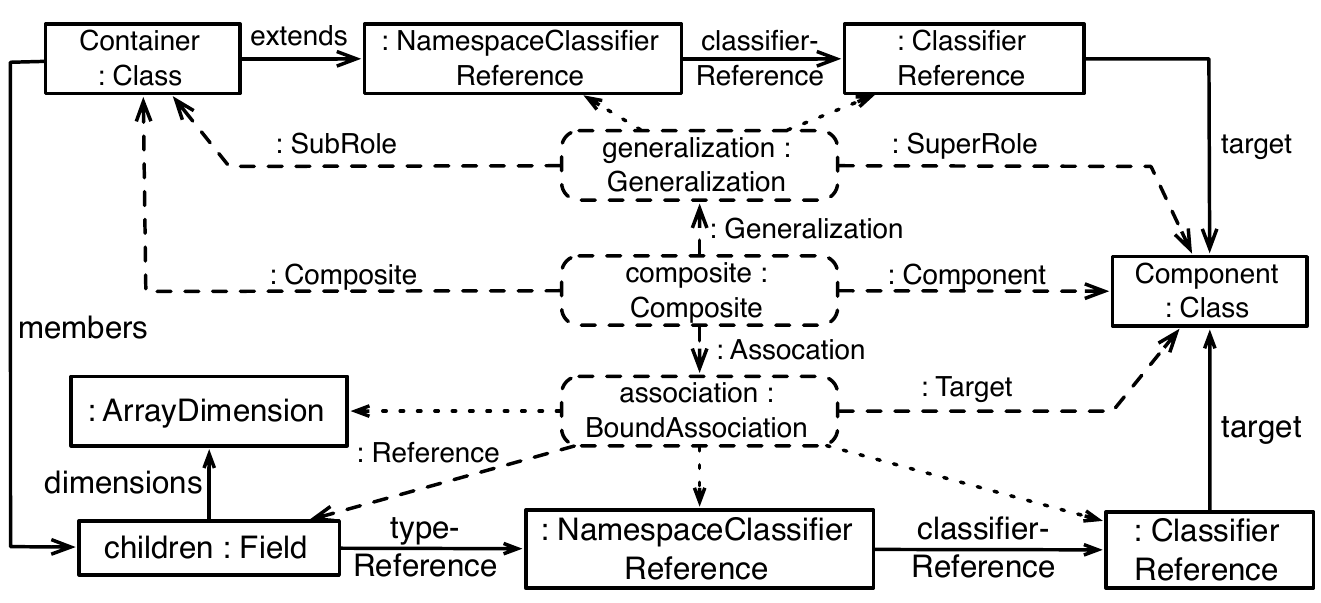}
		\caption{Marked graph pattern matches as UML object model}
		\label{fig:view_instance_graph}
	\end{subfigure}
	\vspace{-2mm}
	\caption{View reference graph (left) and instantiated view graph (right)}
	\vspace{-4mm}
	\label{fig:view_graph}
\end{figure}

According to our running example, Fig. \ref{fig:view_reference_graph} shows an excerpt of a type graph as UML class model that describes which kinds of nodes and edges exist in base graphs and view graphs.
For example, the type graph describes that nodes of type \figemph{Class}, \figemph{Interface} and \figemph{Field} exist in base graphs as denoted by the white classes and that nodes of type \figemph{Generalization}, \figemph{Association}, and \figemph{Composite} exist in view graphs as denoted by gray classes.
The type graph describes which edge types are used in view graphs to describe the roles of nodes in graph pattern matches.
For example, the \figemph{Generalization} node type owns the \figemph{SubRole} and \figemph{SuperRole} edge types to describe that \figemph{Class}  nodes are marked as super- and subclasses in matches of the \figemph{Generalization} graph pattern.
%
%
%
\figabbrvref{\ref{fig:view_instance_graph}} shows a view graph as UML object model that is an instance of the type graph in \figabbrvref{\ref{fig:view_reference_graph}}.
%
%
Solid rectangles denote typed nodes of base graphs.
Solid lines denote edges between nodes of base graphs.
Dashed rounded rectangles denote typed nodes in view graphs, which represent graph pattern matches.
Dashed lines denote typed edges in view graphs, which describe the roles of nodes in marked graph pattern matches.
Dotted lines denote scopes, which are edges that mark nodes without explicit roles in graph pattern matches.
\figabbrvref{\ref{fig:view_instance_graph}} shows a \figemph{Generalization} node that marks a graph pattern match for the generalization graph pattern.
%
The \figemph{SubRole} and \figemph{SuperRole} edges owned by the \figemph{Generalization} node describe that the container class acts as subclass and the component class acts as superclass.
The \figemph{Generalization} node references the namespace and classifier reference via scopes, because both nodes belong to the graph pattern match as well.
\figabbrvref{\ref{fig:view_instance_graph}} depicts a \figemph{BoundedAssociation} node that marks a graph pattern match for the bounded association graph pattern.
%
The \figemph{Reference} and \figemph{Target} edges owned by the \figemph{BoundedAssociation} node describe that the children field acts as reference and the component class acts as target type of the reference.
The \figemph{BoundedAssociation} node owns scopes that reference the array dimension, namespace, and classifier reference, because they belong to the graph pattern match as well.
\figabbrvref{\ref{fig:view_instance_graph}} shows a \figemph{Composite} node that marks a graph pattern match for the \figemph{Composite} graph pattern.
%
The \figemph{Component} and \figemph{Composite} edges owned by the \figemph{Composite} node describe that the component class acts as component and the container class acts as composite of the detected Composite design pattern.
The \figemph{Generalization}  and \figemph{Association} edges owned by the \figemph{Composite} node mark the reused matches of the \figemph{Generalization} and \figemph{Association} graph patterns.
%
%

\subsection{Definition of View Graphs}
\label{sec:view_definition}
%
\structure{Review 2c}
Our approach enables to specify view modules that encapsulate graph transformation rules, which search and mark graph pattern matches by creating nodes and edges in view graphs.
%
%
%
Each view module owns input connectors that describe which kinds of nodes are required by the hidden graph transformation rule.
Each view module owns an output connector to describe which kind of node is created in view graphs by the hidden graph transformation rule to mark graph pattern matches.
%
\structure{Review 2c}
Our approach is independent from graph transformation languages, because view modules hide graph transformation rules.

\begin{figure}[tbp]
	\centering
	\includegraphics[width=0.9\textwidth]{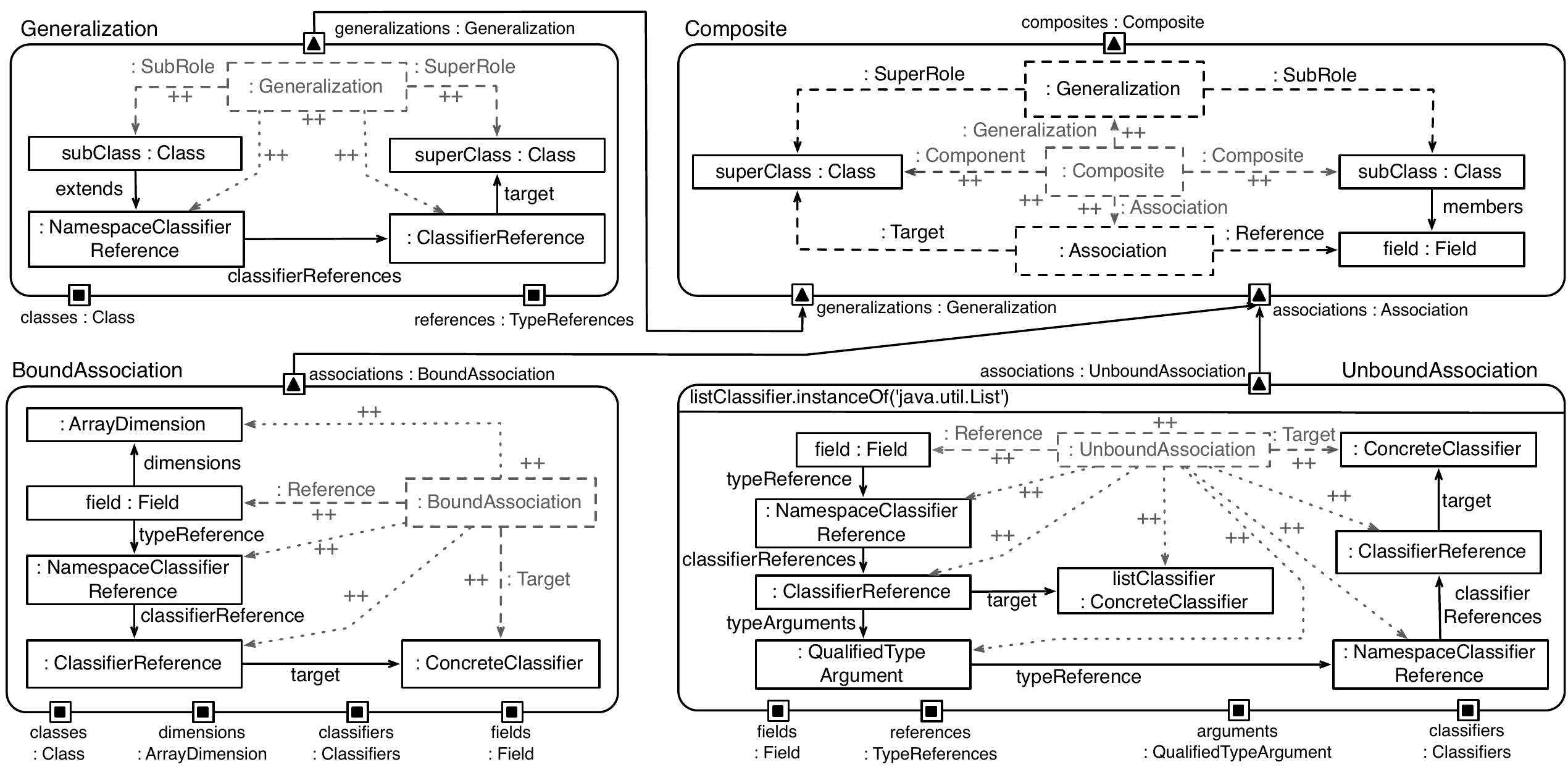}
	\vspace{-2mm}
	\caption{Generalized discrimination network of view modules}
	\vspace{-6mm}
	\label{fig:view_definition}
\end{figure}

\figabbrvref{\ref{fig:view_definition}} shows a generalized discrimination network that consists of the view modules \figemph{Generalization}, \figemph{BoundedAssociation}, \figemph{UnboundedAssociation}, and \figemph{Composite}.
The \figemph{Generalization} view module describes that \figemph{Class} nodes and \figemph{Type\-Reference} (super type of \figemph{Namespace\-Classifier\-Reference} and \figemph{Classifier\-Reference}) nodes are required to produce nodes that mark matches of the \figemph{Generalization} graph pattern.
The view modules \figemph{BoundedAssociation} and \figemph{UnboundedAssociation} create nodes of type \figemph{Association} that mark matches of the \figemph{BoundedAssociation} and \figemph{UnboundedAssociation} graph pattern.
The \figemph{Composite} view module requires nodes of type \figemph{Generali\-zation} and \figemph{Association} to create nodes in view graphs that mark matches of the \figemph{Composite} graph pattern.

\structure{Review 3e}
\figabbrvref{\ref{fig:view_definition}} shows view modules that implement graph transformation rules for creating and maintaining view graphs that enumerate matches for generalizations, association, and composite graph patterns.
When a graph transformation rule finds a graph pattern match, it marks the match by creating a node of a certain node type, edges of certain edge types, and scopes in view graphs to mark which nodes satisfy the graph pattern.
%
%
\figabbrvref{\ref{fig:view_definition}} depicts graph pattern nodes as solid and dashed rectangles.
\figabbrvref{\ref{fig:view_definition}} depicts graph pattern edges as solid and dashed lines.
Solid rectangles and lines refer to nodes and edges in base graphs.
Dashed rectangles and lines refer to nodes and edges in view graphs.
Graph pattern nodes and edges can consist of create modifiers that are depicted as ``++'' (adapted from story diagrams \cite{Fischer:2000ut}).
These create modifiers describe which nodes and edges are created in view graphs when matches for graph patterns are found.
Nodes and edges without create modifier depict the left-hand side of the graph transformation rule.
Nodes and edges with and without create modifier depict the right-hand side of the graph transformation rule.

The graph patterns described in \figabbrvref{\ref{fig:running_example}} define the left-hand side of the graph transformation rules depicted by \figabbrvref{\ref{fig:view_definition}}.
The right-hand side of the depicted graph transformation rules describe which kinds of node and edges are created to mark graph pattern matches.
The graph transformation rules mark all nodes of graph pattern matches.
Nodes with special roles in the graph pattern match are marked by edges with certain edge types (\cf{} dashed lines).
Nodes without special roles in graph pattern matched are marked by scopes (\cf{} dotted lines).
For example, the \figemph{Generalization} view module marks with \figemph{SuperRole} and \figemph{SubRole} edges which class acts as super- and sub\-class.
The \figemph{BoundedAssociation} and \figemph{UnboundedAssociation} view modules marks with \figemph{Reference} and \figemph{Target} edges which field acts as reference and which classifier is the target of the reference.
The \figemph{Composite} view module reuses nodes that mark matches for the \figemph{Generalization} and \figemph{Association} graph patterns.
The \figemph{Composite} view module marks with \figemph{Composite}, \figemph{Component}, \figemph{Generalization}, and \figemph{Association} edges which class acts as composite and component and which generalization and association graph pattern matches are reused.

\subsubsection{Mapping Graph Conditions}
\label{sec:discussion_nested_conditions}
\begin{figure}[tb]
	\centering
	\begin{subfigure}[b]{.19\linewidth}
		\centering
		\includegraphics[scale=0.65]{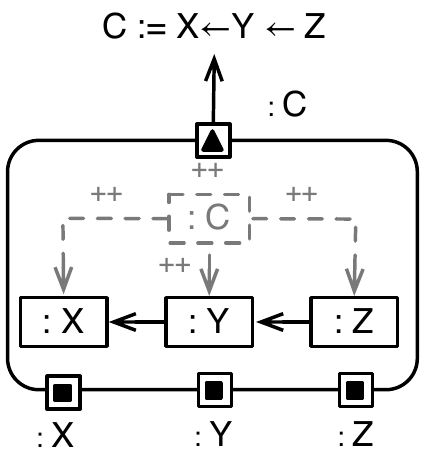}
		\caption{Atomic condition}
		\label{fig:view_mapping_atomic}
	\end{subfigure}
	\begin{subfigure}[b]{.16\linewidth}
		\centering
		\includegraphics[scale=0.65]{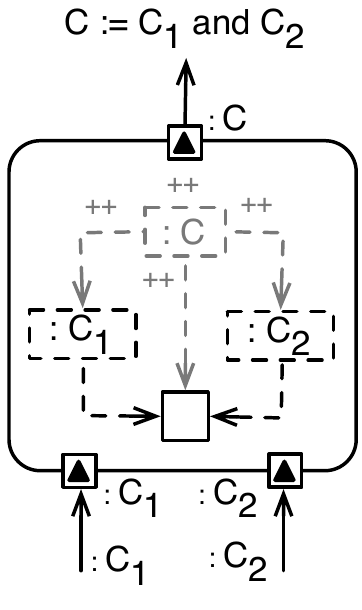}
		\caption{Conjunction}
		\label{fig:view_mapping_conjunction}
	\end{subfigure}
	\begin{subfigure}[b]{.16\linewidth}
		\centering
		\includegraphics[scale=0.65]{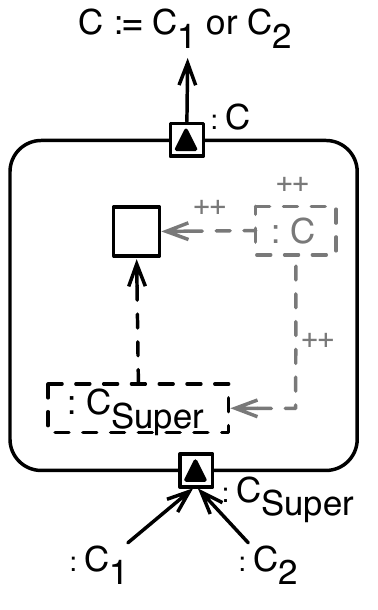}
		\caption{Disjunction}
		\label{fig:view_mapping_disjunction}
	\end{subfigure}
	\begin{subfigure}[b]{.22\linewidth}
		\centering
		\includegraphics[scale=0.65]{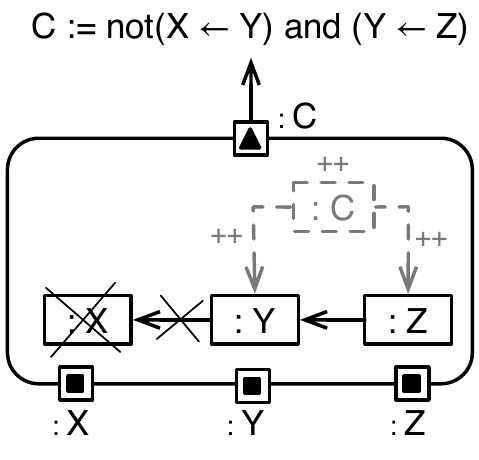}
		\caption{Simple NAC}
		\label{fig:view_mapping_simple_nac}
	\end{subfigure}
	\begin{subfigure}[b]{.22\linewidth}
		\centering
		\includegraphics[scale=0.65]{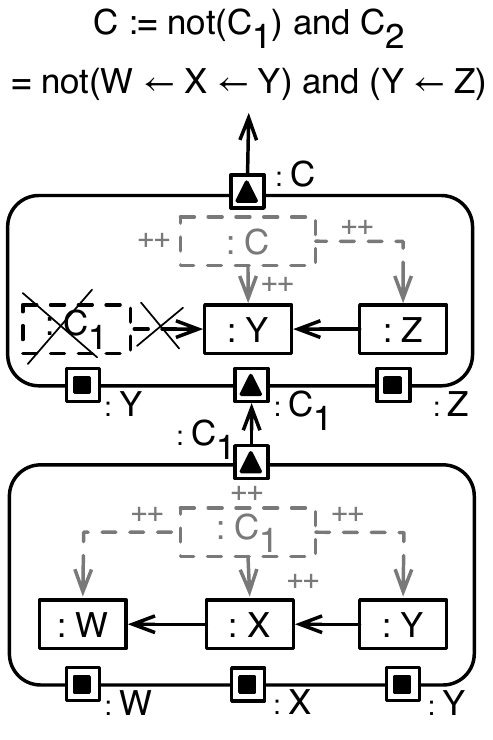}
		\caption{Complex NAC}
		\label{fig:view_mapping_complex_nac}
	\end{subfigure}
	\vspace{-3mm}
	\caption{Schematic mapping of graph conditions to a discrimination network of view modules}
	\vspace{-6mm}
	\label{fig:nested_graph_conditions}
\end{figure}

%
\figabbrvref{\ref{fig:nested_graph_conditions}} shows a schematic mapping of graph conditions \cite{Ehrig:2004aa} to our network of view modules.
If $c$ is a graph condition, then also $\neg c$ is a graph condition.
Furthermore, if $c_{i}$ is a graph condition, then $\vee c_{i}$ and $\wedge c_{i}$ with index set $i \in I$ are graph conditions.

\inlineparagraph{Atomic Graph Condition}
Atomic graph conditions (\cf{} \figabbrvref{\ref{fig:view_mapping_atomic}}) are mapped to single view modules that only receive nodes of base graphs.
Atomic graph conditions enable to express basic conditions on graphs, \eg{}, the existence of certain nodes and edges.
According to our running example, the \figemph{Generalization}, \figemph{BoundedAssociation}, and \figemph{UnboundedAssociation} view modules implement atomic graph conditions.

\inlineparagraph{Conjunction}
View modules with more than one input connector that receive nodes of view graphs implement conjunctions (\cf{} \figabbrvref{\ref{fig:view_mapping_conjunction}}).
According to our running example, the \figemph{Composite} view module implements a conjunction of graph conditions for generalizations and associations.

\inlineparagraph{Disjunction}
View modules with an input connector, which receives nodes from more than one predecessor view module, implement disjunctions (\cf{} \figabbrvref{\ref{fig:view_mapping_disjunction}}).
The input connector must specify the node super type of required nodes of view graphs.
According to our running example, the \figemph{Composite} view module implements a disjunction of graph conditions for unbounded and bounded associations.
Therefore, the \figemph{Composite} view module consists of an input connector with \figemph{Association} node type.

%
\inlineparagraph{Simple NAC}
We refer to the term simple negative application conditions (NAC), when all negated graph patterns nodes are directly connected to non-negated graph pattern nodes, \ie{}, positive application conditions (PACs).
Our approach maps simple NACs to negated graph pattern nodes that refer to base graphs within graph transformation rules of view modules (\cf{} \figabbrvref{\ref{fig:view_mapping_simple_nac}}).

\begin{figure}[tb]
	\centering
	\begin{subfigure}[b]{.49\linewidth}
		\centering
		\includegraphics[scale=0.55]{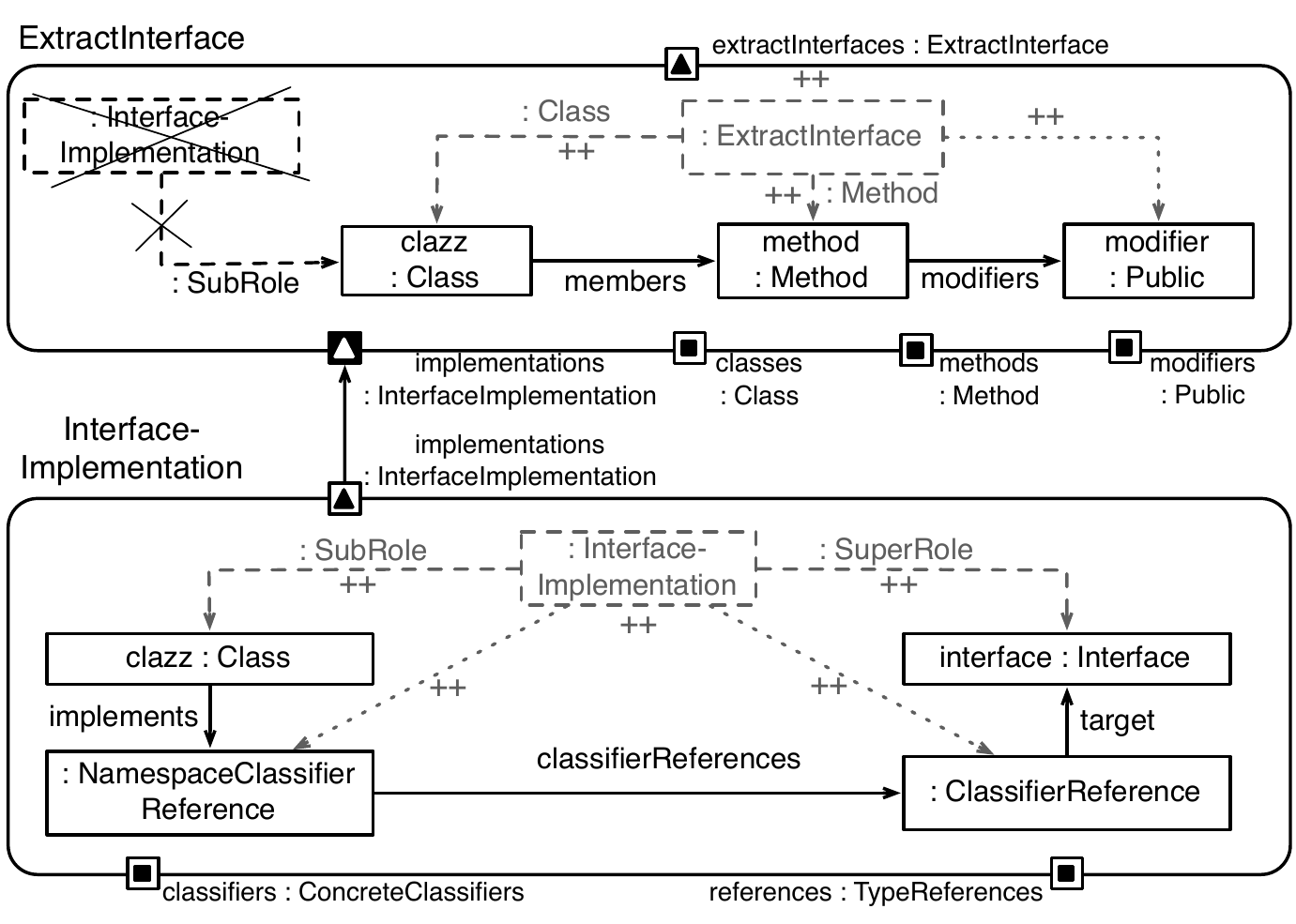}
		 \vspace{-2mm}
		\caption{Complex NAC for \figemph{Extract Interface}}
		\label{fig:complex_nac}
	\end{subfigure}
	\begin{subfigure}[b]{.49\linewidth}
		\centering
		\includegraphics[scale=0.58]{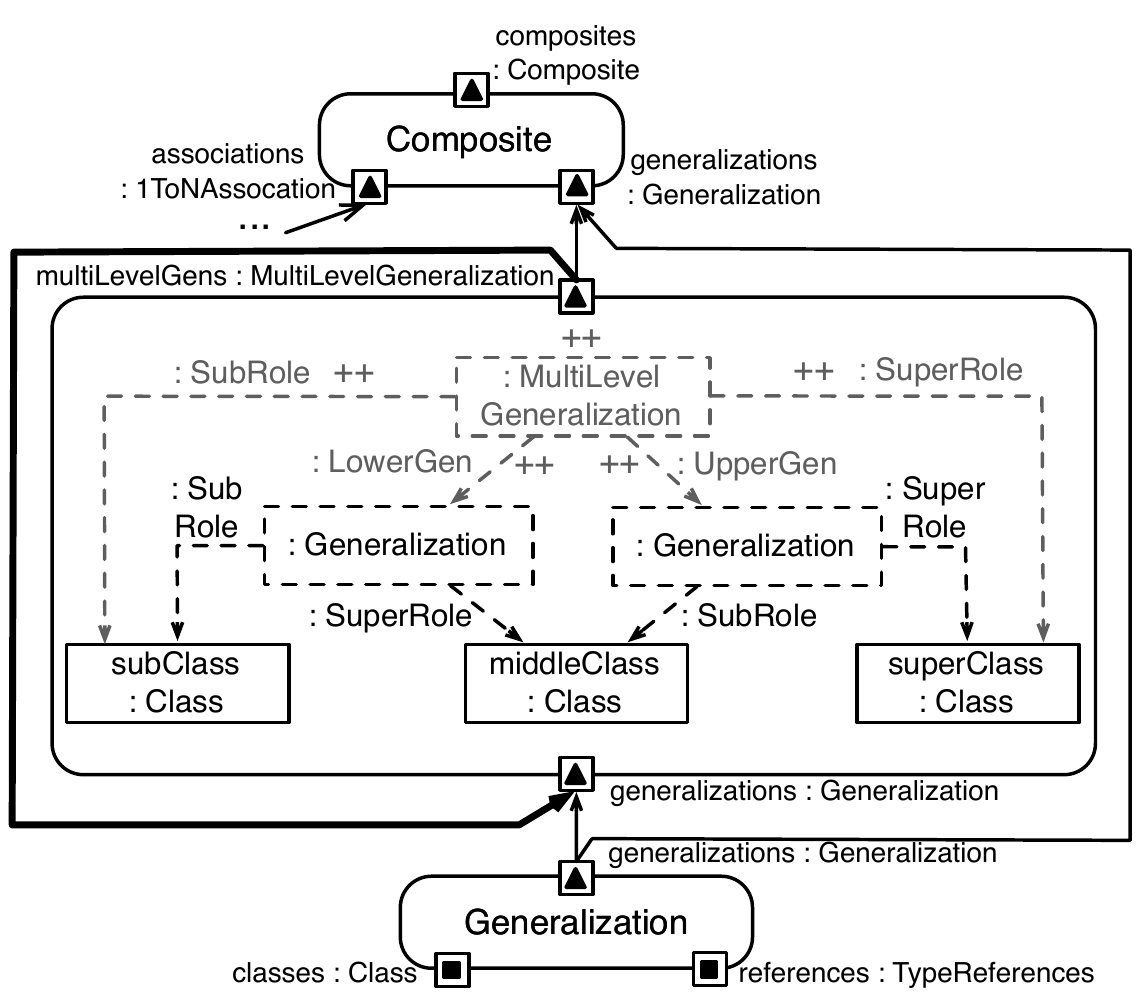}
		 \vspace{-2mm}
		\caption{Recursive definition of \figemph{Multi-Level Generalization}}
	    \label{fig:view_definition_recursion}
    \end{subfigure}
    \vspace{-3mm}
	\caption{Complex NAC and recursion}
	\vspace{-6mm}
	\label{fig:nac_and_recursion}
\end{figure}

\inlineparagraph{Complex NAC}
We refer to the term complex NAC, when at least one negated graph pattern node is not \textit{directly} connected to a PAC.
Our approach splits up complex NACs in two view modules (\cf{} \figabbrvref{\ref{fig:view_mapping_complex_nac}}).
The first view module searches for graph pattern matches for the negated part of the graph condition without negation.
For example, instead of searching for graph pattern matches that satisfy graph condition $\neg c$, the first view module searches for graph pattern matches that satisfy graph condition $c$ and creates nodes in view graphs that mark theses graph pattern matches, accordingly.
Then, the second view module checks that no node in view graphs exists that satisfies graph condition $c$.
\figabbrvref{\ref{fig:complex_nac}} shows an example for mapping complex NACs to view modules.
\figabbrvref{\ref{fig:complex_nac}} shows two view modules that implement graph patterns for \figemph{InterfaceImplementation} and \figemph{ExtractInterface}.
The \figemph{InterfaceImplementation} view module marks graph pattern matches for classes that implement an interface.
The \figemph{ExtractInterface} view module marks graph pattern matches for classes that own public methods, but do \textit{not} implement interfaces.
The latter is denoted by the crossed out \figemph{InterfaceImplementation} node and \figemph{SubRole} edge.
%

\subsubsection{Recursion}
%
Recursion is mapped to cyclic dependencies of view modules, \eg{}, to search for graph patterns that employ path expressions.
These cycles can consist of multiple view modules.
In general, one view module that does not belong to the cycle itself describes the recursion start, while view modules within the cycle describe the recursion step.
According to our running example, the \figemph{Composite} view module has to consider multi-level generalizations as well to also detect variants of the Composite design pattern that employ multiple inheritance levels.
\figabbrvref{\ref{fig:view_definition_recursion}} shows a recursive definition to find graph patterns matches for multi-level generalizations.
The \figemph{Generalization} view module (\cf{} \figabbrvref{\ref{fig:view_definition}}) describes the recursion start.
The \figemph{Multi\-Level\-Generalization} view module in \figabbrvref{\ref{fig:view_definition_recursion}} describes the recursion step, because 
nodes of view graphs that mark graph pattern matches for multi-level generalizations can lead to additional matches for multi-level generalizations.
%
Therefore,  the \figemph{Multi\-Level\-Generalization} view module consists of a dependency between its own output and input connector (\cf{} bold line in \figabbrvref{\ref{fig:view_definition_recursion}}).
\figabbrvref{\ref{fig:view_reference_graph}} shows that the \figemph{MultiLevelGeneralization} node type is a specialization of the \figemph{Generalization} node type and, additionally, owns the \figemph{LowerGen} and \figemph{UpperGen} edge types that describe which reused nodes of type \figemph{Generalization} act as lower and upper generalization.
Note that lower and upper generalizations can be multi-level generalizations as well.
\figabbrvref{\ref{fig:view_definition_recursion}} shows the implementation of the \figemph{Multi\-Level\-Generalization} view module that checks whether two (multi-level) generalizations exist that have a class in common, which acts as super\-class in one generalization and as sub\-class in another generalization.
If yes, the \figemph{Multi\-Level\-Generalization} view module creates a node that marks both \figemph{Generalization} nodes and the super- and sub\-class of the detected multi-level generalization.
%
%

\section{Incremental Maintenance of Graph Database Views}
\label{sec:algorithm}
In this section, we describe our maintenance algorithm for nodes of view graphs.
Our incremental maintenance algorithm consists of maintenance phases that compute the search space for the execution of view modules.
Each maintenance phase executes view module in analogous modes to a) create nodes of view graphs for new graph pattern matches, b) update existing nodes of view graphs that already mark graph pattern matches, and c) delete nodes of view graphs that do not mark graph pattern matches anymore.
We describe our incremental maintenance algorithm bottom-up.
\secabbrvref{\ref{sec:view_module_operationalization}} describes the view module execution modes.
\secabbrvref{\ref{sec:maintenance_events}} describes how the search space for view modules is computed.
\secabbrvref{\ref{sec:maintenance_phases}} describes the maintenance phases that execute view modules with the computed search space.
%

\subsection{View Module Execution Modes}
\label{sec:view_module_operationalization}
A view module can be executed in three different modes called \figemph{Update}, \figemph{Delete}, and \figemph{Create}.
%
\structure{Overview}
In \figemph{Create} mode, view modules search for graph pattern matches, mark these matches with the help of nodes and edges in view graphs, and return created nodes of view graphs.
In \figemph{Update} mode, view modules check whether nodes of view graphs still mark matches for satisfied graph patterns, set nodes of view graphs obsolete, if they do not mark matches for satisfied graph patterns anymore, and return revised nodes of view graphs.
In \figemph{Delete} mode, view modules delete nodes of view graphs that were set obsolete during \figemph{Update} mode or consist of dangling edges due to deleted nodes of base graphs and view graphs.

\begin{figure}[tbp]
	\centering
	\begin{subfigure}[b]{.2\linewidth}
		\centering
		\includegraphics[scale=0.6]{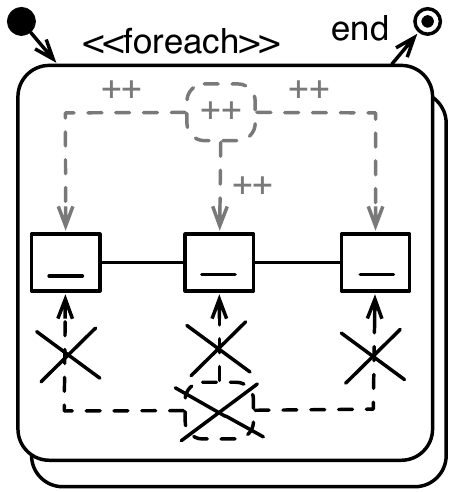}
		\caption{Create mode}
		\label{fig:view_module_operationalization_create}
	\end{subfigure}
	\begin{subfigure}[b]{.2\linewidth}
		\centering
		\includegraphics[scale=0.6]{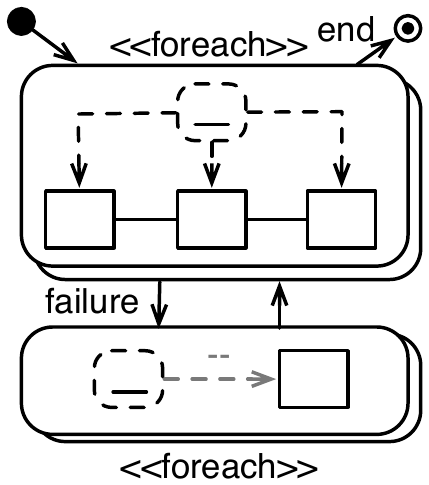}
		\caption{Update mode}
		\label{fig:view_module_operationalization_update}
	\end{subfigure}
	\begin{subfigure}[b]{.2\linewidth}
		\centering
		\includegraphics[scale=0.6]{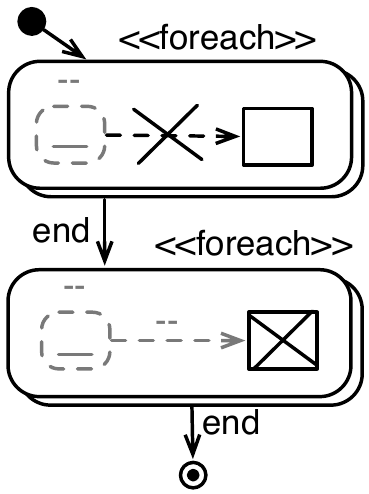}
		\caption{Delete mode}
		\label{fig:view_module_operationalization_delete}
	\end{subfigure}
	\begin{subfigure}[b]{.2\linewidth}
		\centering
		\includegraphics[scale=0.6]{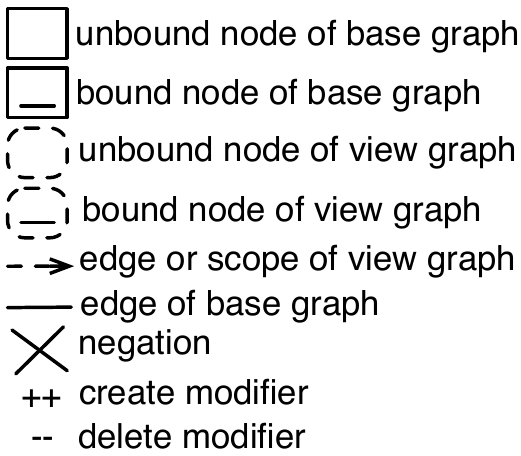}
		\caption{Legend}
		\label{fig:view_module_operationalization_legend}
	\end{subfigure}
	\vspace{-4mm}
	\caption{Schematic graph transformation rules for execution modes of view modules}
	\vspace{-6mm}
	\label{fig:view_module_operationalization}
\end{figure}

\structure{Hidden GT rules}
\figabbrvref{\ref{fig:view_module_operationalization}} shows graph transformation rules hidden by view modules in terms of schematic story diagrams \cite{Fischer:2000ut} to describe the behavior of view modules in each mode.
Each view module that initially created a node of view graphs is responsible for its maintenance.
Nodes of view graphs know which view module created them and view modules know which nodes of view graphs they created.

\inlineparagraph{Create Mode}
In \figemph{Create} mode, view modules receive nodes of base graphs and view graphs according to their input connectors.
These nodes are bound in the graph transformation rule as depicted by \figabbrvref{\ref{fig:view_module_operationalization_create}}.
If these nodes satisfy the left-hand side of the graph transformation rule and the found graph pattern match is \textit{not} already marked by a node of the same type in view graphs as depicted by the negated nodes, edges, and scopes, the graph transformation rule creates a node, edges, and scopes in view graphs that mark all nodes of the found graph pattern match as depicted by the node, edges, and scopes with create modifier.
Thus, view modules mark graph pattern matches at most once.

\inlineparagraph{Update Mode}
In \figemph{Update} mode, view modules receive nodes of view graphs that must be revised to check whether they still mark matches for satisfied graph patterns.
These nodes are bound in the graph transformation rule as depicted by \figabbrvref{\ref{fig:view_module_operationalization_update}}.
If these nodes of view graphs do not mark matches for satisfied graph patterns anymore, the graph transformation rule sets these nodes obsolete as depicted by the \figemph{failure} edge and removal of all edges that are used to mark the graph pattern match.
Otherwise, the node, edges, and scopes of view graphs are preserved.

\inlineparagraph{Delete Mode}
In \figemph{Delete} mode, view modules receive nodes of view graphs that are obsolete.
These nodes are bound in the graph transformation rule as depicted by \figabbrvref{\ref{fig:view_module_operationalization_delete}}.
%
%
The graph transformation rule deletes nodes and their edges from view graphs, if they do not consist of edges or scopes anymore (\cf{} activity on top) or consist of dandling edges or scopes that do \textit{not} mark a node anymore (\cf{} activity at the bottom).

\subsection{Computation of View Module Input}
\label{sec:maintenance_events}
Base and view graph changes are used to derive suspicious, obsolete, and missing nodes of view graphs.

\inlineparagraph{Suspicious Nodes}
Nodes of view graphs are suspicious when they are connected to at least one modified node of base graphs, are connected to another suspicious node of view graphs, or view modules created new nodes in view graphs that dissatisfy complex NACs.
A node is modified when an attribute value of the node changed or two nodes are modified when an edge is added or deleted that connects both nodes.
Our approach uses modified nodes of base graphs to look up connected suspicious nodes of view graphs.
Furthermore, when view modules create new nodes in view graphs, complex NACs implemented by dependent view modules may become dissatisfied.
Therefore, our approach employs a reachability test to collect suspicious nodes of view graphs, when view modules create new nodes.
The reachability test collects nodes of view graphs that are directly or indirectly reachable from created nodes in view graphs.
The reachability test only traverses nodes, if they have the same node \mbox{(sub-)type} as input connectors of view modules that dependent on the view module that created new nodes.
For example, when the \figemph{InterfaceImplementation} view module (\cf{} \figabbrvref{\ref{fig:complex_nac}}) creates a new node of type \figemph{InterfaceImplementation} in view graphs, the reachability test looks up all reachable nodes of type \figemph{ExtractInterface} by traversing nodes of type \figemph{Class}, \figemph{Method}, and \figemph{Public} (\cf{} \figabbrvref{\ref{fig:complex_nac}}).
In \figemph{Update} mode, view modules use suspicious nodes of view graphs to check whether they still mark matches for satisfied graph patterns.

\inlineparagraph{Obsolete Nodes}
A node of view graphs is \textit{obsolete} when the node consists of at least one dangling edge or scope that is not connected to a node anymore.
Our approach uses deleted nodes of base graphs to look up nodes of view graphs that became obsolete.
In \figemph{Delete} mode, view modules delete obsolete nodes.

\inlineparagraph{Missing Nodes}
A \textit{missing} node of view graphs is a node that currently does not exist in view graphs, although it must exist due to changes of base graphs that result in new graph pattern matches.
Our approach employs a reachability test that collects nodes of base graphs and view graphs that may result in new graph pattern matches.
%
%
The reachability test collects all nodes that are directly or indirectly reachable from a) created and modified nodes of base graphs or b) are marked by created / were marked by deleted nodes of view graphs.
The reachability test only collects nodes, if they have the same node \mbox{(sub-)type} as input connectors of the view module.
For example, when a node of type \figemph{Class} is added to base graphs, the reachability test for the \figemph{Generalization} view module (\cf{} \figabbrvref{\ref{fig:view_definition}}) collects all nodes of type \figemph{Class} and \figemph{TypeReference} that are reachable from the added node.
%
%
%
%
We assume that view modules employ graph patterns that are connected graphs and edges between nodes can be traversed bidirectionally.


\subsection{Maintenance Phases}
\label{sec:maintenance_phases}
\begin{wrapfigure}{r}{6.3cm}
	{
		\vspace{-5mm}
		\scriptsize
		\begin{algorithmic}[0]
			\Procedure{maintain}{events}
			\State suspicious := $\emptyset$
			\Repeat
			\State suspicious:= suspicious $\cup$ suspiciousNodes(events)
			\State obsoletes := \Call{update}{suspicious}
			\State obsoletes := obsoletes $\cup$ obsoletesNodes(events)
			\State changed := \Call{delete}{obsoletes}
			\State changed := changed $\cup$ changedNodes(events)
			\State suspicious := \Call{create}{changed}
			\State events := $\emptyset$
			\Until{suspicious = $\emptyset$}
			\EndProcedure
		\end{algorithmic}
		\vspace{-3mm}
	}
	\caption{Order of maintenance phases}
	\vspace{-3mm}
	\label{algo:phases}
\end{wrapfigure}

Our maintenance algorithm employs the subsequent maintenance phases \figemph{Update}, \figemph{Delete}, and \figemph{Create} to maintain suspicious, obsolete, and missing nodes of view graphs.
\figabbrvref{\ref{algo:phases}} shows the order of the maintenance phases.
The algorithm passes suspicious nodes of view graphs to the \figemph{Update} phase, obsolete nodes of view graphs to the \figemph{Delete} phase, and added / modified nodes of base and view graphs to the \figemph{Create} phase.
If the \figemph{Create} phase returns new suspicious nodes of view graphs, the algorithm executes an additional cycle of \figemph{Update}, \figemph{Delete}, and \figemph{Create} phases to revise these suspicious nodes.

\inlineparagraph{Update Phase}
\figabbrvref{\ref{algo:phases_update}} shows pseudo code for the \figemph{Update} phase.
For each suspicious node of view graphs, the algorithm looks up the responsible view module, passes the suspicious node to the view module, and executes the view module in \figemph{Update} mode.
%
If the view module sets the node obsolete, the node is added to the set of obsolete nodes.
Otherwise, all nodes that depend on the updated node are updated as well.
Finally, the \figemph{Update} phase returns all collected obsolete nodes.

\inlineparagraph{Delete Phase}
\figabbrvref{\ref{algo:phases_delete}} shows pseudo code for the \figemph{Delete} phase.
For each obsolete node of view graphs, the algorithm looks up the responsible view module, passes the obsolete node to the view module, and executes the view module in \figemph{Delete} mode.
%
The view module returns all nodes that were previously marked by the deleted obsolete node.
The algorithm considers the returned nodes as changed and collects them.
Afterwards, all nodes that depend on the deleted node are removed as well.
Finally, the algorithm returns all nodes that were previously marked by the deleted obsolete nodes of view graphs.

\inlineparagraph{Create Phase}
\figabbrvref{\ref{algo:phases_create}} shows pseudo code for the \figemph{Create} phase.
The algorithm iterates the network of view modules with respect to recursion cycles.
The algorithm executes each view module in \figemph{Create} mode and passes candidate nodes, which may lead to new graph pattern matches (\cf{} \secabbrvref{\ref{sec:maintenance_events}}), to view modules.
The view module returns all created nodes of view graphs that mark new graph pattern matches.
The algorithm uses the created nodes to determine nodes of view graphs that become suspicious (\cf{} \secabbrvref{\ref{sec:maintenance_events}}).
These suspicious nodes are collected and returned at the end of the \figemph{Create} phase.

\begin{figure}[tbp]
	\centering
	\begin{subfigure}[b]{.29\linewidth}
		\centering
		{
			\tiny
			\begin{algorithmic}[0]
				\Procedure{update}{suspiciousNodes}
				\State obsoletes := $\emptyset$
				\For{node in suspiciousNodes}
					\State module := node.module
					\State module.update(node)
					\If{node is obsolete}
						\State obsoletes := obsoletes $\cup$ \{node\}
					\Else
						\State dependents := node.dependents
						\State obsoletes := obsoletes $\cup$ \Call{update}{dependents}
					\EndIf
				\EndFor
				\State \Return obsoletes
				\EndProcedure
			\end{algorithmic}
		}
		\vspace{-2mm}
		\caption{Update phase}
		\label{algo:phases_update}
	\end{subfigure}
	\begin{subfigure}[b]{.3\linewidth}
		\centering
		\vspace{2mm}
		{
			\tiny
			\begin{algorithmic}[0]
				\Procedure{delete}{obsoleteNodes}
				\State changed := $\emptyset$
				\For{node in obsoleteNodes}
					\State module := node.module
					\State module.delete(node)
					\State changed := changed $\cup$ \{previously marked nodes\}
					\State dependents := node.dependents
					\State changed := changed $\cup$ \Call{delete}{dependents}
				\EndFor
				\State \Return changed
				\EndProcedure
			\end{algorithmic}
		}
		\vspace{-2mm}
		\caption{Delete phase}
		\label{algo:phases_delete}
	\end{subfigure}
	\begin{subfigure}[b]{.39\linewidth}
		\centering
		\vspace{2mm}
		{
			\tiny
			\begin{algorithmic}[0]
				\Procedure{create}{changedNodes}
				\State suspicious := $\emptyset$
				\State result := $\emptyset$
				\While{hasNextModule(result)} \Comment{handles recursion}
					\State module := nextModule(result) \Comment{handles recursion}
					\State candidates := reachabilityMissing(changedNodes,module)
					\State result := module.create(candidates)
					\State dependents := module.dependents
					\State suspicious := suspicious $\cup$ reachabilitySuspicious(result,dependents)
				\EndWhile
				\State \Return suspicious
				\EndProcedure
			\end{algorithmic}
		}
		\vspace{-2mm}
		\caption{Create phase}
		\label{algo:phases_create}
	\end{subfigure}
	\vspace{-3mm}
	\caption{Execution of view module during maintenance phases}
	\vspace{-7mm}
	\label{algo:pseudo_code}
\end{figure}

\subsubsection{Positive and Negative Application Conditions}
\label{sec:maintenance_pac}
Graph conditions must be mapped to our view modules as described in \secabbrvref{\ref{sec:discussion_nested_conditions}}.

\inlineparagraph{PACs}
Created and modified nodes of base and view graphs may satisfy PACs.
Deleted and modified nodes of base and view graphs may dissatisfy PACs.
%
%
The \figemph{Update} phase derives suspicious nodes of view graphs from modified nodes (\cf{} \secabbrvref{\ref{sec:maintenance_events}}) and sets these suspicious nodes obsolete, if required.
The \figemph{Delete} phase deletes obsolete nodes of view graphs that were set obsolete by the previous \figemph{Update} phase or are obsolete due to deleted nodes of base and view graphs.
%
The \figemph{Create} phase uses created and modified nodes of base graphs as well as  nodes of view graphs created and deleted by predecessor view modules to compute the candidate nodes for view modules.
Thus, our algorithm supports PACs.

\inlineparagraph{Simple NACs}
Deleted and modified nodes of base graphs may satisfy simple NACs.
Nodes in base graphs that were connected to a deleted node of base graphs are considered as modified.
Created and modified nodes of base graphs are used to compute the candidate nodes for view modules in the \figemph{Create} phase (\cf{} \secabbrvref{\ref{sec:maintenance_events}}).
Thus, the \figemph{Create} phase detects satisfied simple NACs.
Created and modified nodes of base graphs may dissatisfy simple NACs.
When a node of base graphs is added via an edge to another node of base graphs, both nodes are considered as modified.
Modified nodes are used to derive suspicious nodes in view graphs (\cf{} \secabbrvref{\ref{sec:maintenance_events}}).
Thus, the \figemph{Update} phase detects dissatisfied simple NACs.

\inlineparagraph{Complex NACs}
Created nodes of view graphs may dissatisfy complex NACs and, thus, may make other nodes of view graphs obsolete.
When the \figemph{Create} phase creates new nodes in view graphs, our algorithm looks up suspicious nodes of view graphs (\cf{} \secabbrvref{\ref{sec:maintenance_events}}).
%
If such suspicious nodes exist, an additional \figemph{Update}, \figemph{Delete}, and \figemph{Create} sequence is employed to revise these suspicious nodes.
Thus, dissatisfied complex NACs are detected during the \figemph{Update} phase.
Deleted nodes of view graphs may satisfy complex NACs and, thus, cause missing nodes in view graphs.
Nodes of base and view graphs that were connected to deleted nodes of view graphs are used to compute the candidate nodes for view modules to find missing nodes in view graphs (\cf{} \secabbrvref{\ref{sec:maintenance_events}}).
Thus, satisfied complex NACs are detected in the \figemph{Create} phase.

\subsubsection{Recursion}
\label{sec:maintenance_recursion}

When the discrimination network is acyclic, a topological sorting is sufficient to execute view modules in correct order.
When the discrimination network consists of cyclic dependencies between view modules, an execution plan is generated that considers these cyclic dependencies when sorting view modules for execution.
We refer to cyclic dependencies between view modules as \textit{recursion cycle}.
Recursion cycles of view modules are executed until the fix point view module reaches a fix point.
A fix point view module is a view module that has at least one dependent view module that is \textit{not} part of the recursion cycle or is connected to the termination of the network.
A fix point view module reached a fix point when it did not update, delete, or create nodes of view graphs, because then dependent view modules \textit{within} the recursion cycle are not impacted by the output of the fix point view module anymore.

\structure{Review 1f}
During \figemph{Create} phase, nodes created by a fix point module are passed to dependent view modules \textit{within} the recursion cycle, first.
When the fix point view module reached its fix point, the nodes created by the fix point view module are passed to dependent view modules that do \textit{not} belong to the recursion cycle.
The recursion cycle terminates when the fix point view module has a fix point.
%
%
During \figemph{Create} phase, recursion cycles terminate when the number of nodes in base graphs has an upper bound, nodes of view graphs mark at least one node in base graphs (\ie{}, nodes of view graphs that only mark nodes of view graphs are not permitted), and view modules in the recursion cycle do not create and delete the same kind of nodes in view graphs.
Then, the number of nodes in view graphs has also an upper bound, because each node of base graphs is marked by at most one node of the same type in view graphs due to the NAC of the graph transformation rule in \figemph{Create} mode (\cf{} \figabbrvref{\ref{fig:view_module_operationalization_create}}).
During \figemph{Update} and \figemph{Delete} phase, the edges and scopes between nodes of view graphs are used to revise dependent suspicious nodes and delete dependent obsolete nodes.
Since the number of dependent nodes has an upper bound when the \figemph{Create} phase terminated, the revision and deletion of dependent nodes in view graphs terminates.

\section{Evaluation}
\label{sec:evaluation}
In this section, we evaluate the performance of our incremental maintenance algorithm and compare the performance of ordinary Rete network structures \cite{Forgy:1982cg} with generalized Gator network structures \cite{Hanson:2002aa} for the incremental maintenance of view graphs.
For evaluation purposes, we implemented a batch algorithm as reference algorithm and our incremental maintenance algorithm using the Eclipse Modeling Framework (EMF) and story diagrams \cite{Fischer:2000ut} as operationalization of view modules.
The batch maintenance algorithm \cite{Beyhl:2015tr} considers \textit{all} nodes of view graphs created by a view module as suspicious, checks for \textit{all} nodes of view graphs created by a view module whether they are obsolete, and uses \textit{all} nodes that have a node (sub-)type as specified by input connectors of view modules to find missing nodes in view graphs.
According to our running example, we recover software design patterns in ASGs of Java source code.
We preprocessed the first 100 source code revisions of the Apache Ant and Xerces source code repositories to derive EMF models.
Both algorithms result in exactly the same nodes in view graphs.

First, we compared the performance of the batch and incremental algorithm.
We performed a view maintenance for each revision using either our batch or incremental algorithm.
We used the history of the source code repositories to merge applied modifications into the ASGs using EMFCompare.
We employed 49 view modules \cite{Beyhl:2015tr} to recover design patterns.
We do not consider precision and recall of the recovered design patterns, because they are the same for both algorithms.
Table \ref{tab:evaluation_maintenance} shows the number of nodes in base and view graphs for revision 1 and 100.
For Ant 2,24\% and for Xerces 0,82\% of nodes in base graphs changed between two revisions in average.
The computed candidate sets had a size of 10,36\% for Ant and 4,96\% for Xerces in comparison to the number of nodes passed to view modules during batch maintenance.
In total, the incremental algorithm is approx. 26 times faster for Ant and approx. 53 times faster for Xerces than the batch algorithm.
%

Second, we compared the performance of Rete and Gator network structures.
The runtime performance of a discrimination network depends on the network topology as stated by Varro\etal \cite{Varro:2013bu} for incremental graph pattern matching using Rete networks and Hanson\etal \cite{Hanson:2002aa} for maintaining materialized views of \textit{relational} databases using Gator networks.
For our experiment, we used the Gator network presented in our running example and emulated an equivalent Rete network with 15 modules using our approach.
Finding an optimal network structure is a non-trivial task.
We followed optimization criteria from related work \cite{Bergmann:2008jk,Hanson:2002aa,Varro:2013bu} such as matching to-one reference early in the network and using few network nodes to reduce memory footprint.
We do not use the Rete based matcher \cite{Bergmann:2008jk} of EMF-IncQuery \cite{Bergmann:2010th} to measure the performance of the Rete network, because we aim for comparing concepts instead of different technologies such as different employed graph pattern matchers.
Table \ref{tab:evaluation_comparison} shows in total the time required for incremental maintenance, number of nodes in view graphs, and view graph size for the first 100 revisions.
Our approach is 2,7 times for Ant and 3,1 times for Xerces faster than the equivalent emulated Rete network.
Our approach requires 2,92\% - 8,03\% for Ant and 0,30\% - 0,45\% for Xerces of memory in comparison to the equivalent emulated Rete network.
The measurement proves that Gator networks \textit{can} outperform Rete networks in time and space also for incremental graph pattern matching.

\begin{table}
	\begin{subtable}[b]{1.0\linewidth}
		\centering
		\scriptsize
		\begin{tabular}[htp]{|l||c|c|c|c|c|c|c|c|c|}
			\hline
			\textbf{Projects} & \multicolumn{2}{|c|}{\textbf{\#Nodes in BG}} & \multicolumn{2}{|c|}{\textbf{\#Nodes in VG}} & \multicolumn{2}{|c|}{\textbf{View Size (MB)}} & \multicolumn{2}{|c|}{\textbf{Execution Time for 100 revisions}} & \textbf{Speedup} \\ \hline
			Apache\ldots & Rev.1 & Rev.100 & Rev.1 & Rev.100 & Rev.1 & Rev.100 & Batch & Incremental & \\ \hline \hline
			Ant & 12442 & 22071 & 1725 & 2927 & 1,2 & 2,1 & 0 h 9 min 14 s & 0 min 21 s & 26,38\\ \hline
			Xerces & 133858 & 191246 & 20050 & 25636 & 13,3 & 16,7 & 15 h 33 min 44 s & 17 min 42 s & 52,75\\ \hline
		\end{tabular}
		\vspace{-2mm}
		\caption{Comparison of our batch and maintenance algorithms concerning execution time}.
		\label{tab:evaluation_maintenance}
	\end{subtable}
	\begin{subtable}[b]{1.0\linewidth}
		\centering
		\scriptsize
		\vspace{-1mm}
		\begin{tabular}[htp]{|l||c|c|c|c|c|c|c|}
			\hline
			\textbf{Projects} & \multicolumn{3}{|c|}{\textbf{Emulated Rete}} & \multicolumn{3}{|c|}{\textbf{Our Approach}} & \textbf{Speedup}  \\ \hline
			Apache\ldots & Incremental & \#Nodes in VG & View Size (MB) & Incremental & \#Nodes in VG & View Size (MB) & \\ \hline \hline
			Ant & 4,9 s & 296 - 437 & 0,171 - 0,249 & 1,8 s & 5 - 20 &  0,005 - 0,020 & 2,7 \\ \hline
			Xerces & 43,5 s & 1155 - 1764 & 0,657 - 1,003 & 13,9 s & 2 & 0,003 & 3,1 \\ \hline
		\end{tabular}
		\vspace{-2mm}
		\caption{Comparison of emulated Rete network and our approach concerning memory footprint and execution time}
		\label{tab:evaluation_comparison}
	\end{subtable}
	\vspace{-8mm}
	\caption{Overview of evaluation results (BG: base graph; VG: view graph)}
	\vspace{-6mm}
\end{table}

\section{Related Work}
\label{sec:related_work}
We describe \textit{discrimination networks} in \secabbrvref{\ref{sec:state_of_the_art}}.
No research exists that employs Gator networks \cite{Hanson:2002aa} for incremental graph pattern matching or view maintenance of \textit{graph} databases, although the authors showed that optimized Gator networks \textit{can} outperform Rete networks for view maintenance of \textit{relational} databases.
No research exists, which shows that Gator networks \textit{can} outperform Rete networks for incremental graph pattern matching.
In contrast to existing approaches for incremental graph pattern matching, our approach supports Gator networks with general network structures, while Rete network based approaches are limited to network nodes with at most two inputs.
%
\structure{Review 1e}
Thus, Rete networks require more network nodes to maintain matches for certain graph patterns and, thus, have to store more intermediate graph pattern matches than Gator networks.
Thus, Rete networks require more comparison operations than Gator networks, when they update the state of the network.
Thus, also for incremental graph pattern matching Gator networks \textit{can} outperform Rete networks in space and time.
In contrast to Rete and Gator networks, our approach supports cyclic network structures to enable recursion.

Database view maintenance is often performed incrementally.
\textit{Relational} databases employ an impact analysis \cite{Harrison:1992aa}, derive incremental maintenance queries that transfer views into a consistent state \cite{Qian:1991aa}, or employ discrimination networks for view maintenance \cite{Hanson:2002aa}.
\textit{Object-oriented} databases make use of object-oriented concepts and map objects to tables.
For view maintenance, object-oriented databases re-write queries to make all tables explicit in view definitions to also consider inherited and inheriting tables \cite{Liu:2000aa}.
\textit{Graph} databases such as GRAS \cite{Kiesel:1993aa}, GRACE \cite{Srinivasa:2005ab}, and Neo4j \cite{Robinson:2015aa} do not provide concepts to define and maintain graph database views.
Zhuge\etal \cite{Zhuge:1998aa} introduce the notion of graph-structured databases and employ delegate nodes that reference nodes in graphs to constitute views.
We extended their delegate concept by adding node types to delegate nodes and edge types to effectively refer to nodes with certain roles in graph pattern matches.
However, the authors limit their approach to tree-structured data and employ selection paths and conditions (\ie{}, no graph patterns) to define views.
%
%
%

\textit{Graph indexing} approaches focus on path-based indexing (\eg{} APEX \cite{Chung:2002aa}) for fast evaluation of path expressions and approaches that index frequent graph-structures for fast subgraph isomorphism tests (\eg{} gIndex \cite{Yan:2005aa}).
However, these approaches neither maintain graph pattern matches nor focus on the maintenance of such indexes.
%
\textit{Model search} approaches aim for the efficient retrieval of model elements.
For example, Moogle \cite{Lucredio:2010ew} maps model search to text-based search.
%
The maintenance of search indexes is not in scope of these approaches and they do not index and maintain graph pattern matches.

VIATRA2 \cite{Bergmann:2008jk} provides Rete network based matching for incremental model queries over EMF models in EMF-IncQuery \cite{Bergmann:2010th}, the incremental derivation of features (attributes and references) in EMF models \cite{Rath:2012aa}, live model transformations to propagate changes of source models to target models \cite{Rath:2008ex}, and the incremental derivation of view models \cite{Debreceni:2014aa}.
In contrast to our approach, these approaches are limited to Rete networks.
Model constraint evaluation approaches rewrite constraints based on model changes \cite{Cabot:2006bo} to reduce computational complexity when re-evaluating model constraints or employ model profilers to keep track of model element that are traversed when evaluating model constraints to be aware of which modified model elements demand a re-evaluation of model constraints \cite{Egyed:2006ic}.
In our previous work \cite{Seibel:2010jb}, we maintain traceability links incrementally by employing localization rules to create and delete traceability links without any explicit procedure for updating suspicious traceability links.
In this paper, we extended our previous approach to support general graph pattern matches of arbitrary domains by making use of Gator networks and an explicit \figemph{Update} phase for suspicious graph pattern matches.
Ordinary Rete and Gator networks do not consist of an explicit \figemph{Update} phase and instead map updates to a sequence of delete and create steps.
We extended our previous approach to support NACs during maintenance.

\section{Conclusion and Future Work}
\label{sec:conclusion}
In this paper, we present an enumeration mechanism that enables to mark and maintain graph pattern matches in graph data effectively and efficiently.
We describe how our enumeration mechanism is used to define views over graph data using view modules.
%
\structure{Review 2f}
These view modules hide graph transformation rules and, therefore, the presented concepts can be easily transferred to arbitrary kinds of graph data and graph databases.
Our incremental maintenance algorithm keeps enumerations of graph pattern matches consistent with graph pattern matches in graph data and supports complex NACs and recursion.
Our evaluation shows that our incremental algorithm scales when the size of the graph data increases and that Gator networks \textit{can} outperform Rete networks in time and space for incremental maintenance of graph pattern matches.
As future work, we prove that our approach is as expressive as nested graph conditions.


\begin{thebibliography}{10}
	\providecommand{\bibitemdeclare}[2]{}
	\providecommand{\surnamestart}{}
	\providecommand{\surnameend}{}
	\providecommand{\urlprefix}{Available at }
	\providecommand{\url}[1]{\texttt{#1}}
	\providecommand{\href}[2]{\texttt{#2}}
	\providecommand{\urlalt}[2]{\href{#1}{#2}}
	\providecommand{\doi}[1]{doi:\urlalt{http://dx.doi.org/#1}{#1}}
	\providecommand{\bibinfo}[2]{#2}
	
	\bibitemdeclare{inproceedings}{Bergmann:2010th}
	\bibitem{Bergmann:2010th}
	\bibinfo{author}{G\'{a}bor \surnamestart Bergmann\surnameend},
	\bibinfo{author}{\'{A}kos \surnamestart Horv\'{a}th\surnameend},
	\bibinfo{author}{Istv\'{a}n \surnamestart R\'{a}th\surnameend},
	\bibinfo{author}{D\'{a}niel \surnamestart Varr\'{o}\surnameend},
	\bibinfo{author}{Andr\'{a}s \surnamestart Balogh\surnameend},
	\bibinfo{author}{Zolt\'{a}n \surnamestart Balogh\surnameend} \&
	\bibinfo{author}{Andr\'{a}s \surnamestart \"{O}kr\"{o}s\surnameend}
	(\bibinfo{year}{2010}): \emph{\bibinfo{title}{{Incremental Evaluation of
				Model Queries over EMF Models}}}.
	\newblock In: {\sl \bibinfo{booktitle}{Model Driven Engineering Languages and
			Systems}}, {\sl \bibinfo{series}{LNCS}} \bibinfo{volume}{6394},
	\bibinfo{publisher}{Springer}, pp. \bibinfo{pages}{76--90},
	\doi{10.1007/978-3-642-16145-2\_6}.
	
	\bibitemdeclare{inproceedings}{Bergmann:2008jk}
	\bibitem{Bergmann:2008jk}
	\bibinfo{author}{G\'{a}bor \surnamestart Bergmann\surnameend},
	\bibinfo{author}{Andr\'{a}s \surnamestart \"{O}kr\"{o}s\surnameend},
	\bibinfo{author}{Istv\'{a}n \surnamestart R\'{a}th\surnameend},
	\bibinfo{author}{D\'{a}niel \surnamestart Varr\'{o}\surnameend} \&
	\bibinfo{author}{Gergely \surnamestart Varr\'{o}\surnameend}
	(\bibinfo{year}{2008}): \emph{\bibinfo{title}{{Incremental Pattern Matching
				in the VIATRA Model Transformation System}}}.
	\newblock In: {\sl \bibinfo{booktitle}{Proceedings of the 3$^{rd}$
			International Workshop on Graph and Model Transformations}},
	\bibinfo{series}{GRaMoT '08}, \bibinfo{publisher}{ACM}, pp.
	\bibinfo{pages}{25--32}, \doi{10.1145/1402947.1402953}.
	
	\bibitemdeclare{techreport}{Beyhl:2015tr}
	\bibitem{Beyhl:2015tr}
	\bibinfo{author}{Thomas \surnamestart Beyhl\surnameend} \&
	\bibinfo{author}{Holger \surnamestart Giese\surnameend}
	(\bibinfo{year}{2015}): \emph{\bibinfo{title}{{Efficient and Scalable Graph
				View Maintenance for Deductive Graph Databases based on Generalized
				Discrimination Networks}}}.
	\newblock \bibinfo{type}{Technical Report}, \bibinfo{institution}{Hasso
		Plattner Institute at the University of Potsdam}.
	\newblock
	\urlprefix\url{http://nbn-resolving.de/urn:nbn:de:kobv:517-opus4-79535}.
	
	\bibitemdeclare{incollection}{Bunke:1991cw}
	\bibitem{Bunke:1991cw}
	\bibinfo{author}{H.~\surnamestart Bunke\surnameend},
	\bibinfo{author}{T.~\surnamestart Glauser\surnameend} \&
	\bibinfo{author}{T.-H. \surnamestart Tran\surnameend} (\bibinfo{year}{1991}):
	\emph{\bibinfo{title}{{An efficient implementation of graph grammars based on
				the RETE matching algorithm}}}.
	\newblock In: {\sl \bibinfo{booktitle}{Graph Grammars and Their Application to
			Computer Science}}, {\sl \bibinfo{series}{LNCS}} \bibinfo{volume}{532},
	\bibinfo{publisher}{Springer}, pp. \bibinfo{pages}{174--189},
	\doi{10.1007/BFb0017389}.
	
	\bibitemdeclare{inproceedings}{Cabot:2006bo}
	\bibitem{Cabot:2006bo}
	\bibinfo{author}{Jordi \surnamestart Cabot\surnameend} \&
	\bibinfo{author}{Ernest \surnamestart Teniente\surnameend}
	(\bibinfo{year}{2006}): \emph{\bibinfo{title}{{Incremental Evaluation of OCL
				Constraints}}}.
	\newblock In: {\sl \bibinfo{booktitle}{Advanced Information Systems
			Engineering}}, \bibinfo{publisher}{Springer}, pp. \bibinfo{pages}{81--95},
	\doi{10.1007/11767138\_7}.
	
	\bibitemdeclare{inproceedings}{Chung:2002aa}
	\bibitem{Chung:2002aa}
	\bibinfo{author}{Chin-Wan \surnamestart Chung\surnameend},
	\bibinfo{author}{Jun-Ki \surnamestart Min\surnameend} \&
	\bibinfo{author}{Kyuseok \surnamestart Shim\surnameend}
	(\bibinfo{year}{2002}): \emph{\bibinfo{title}{{APEX: An Adaptive Path Index
				for XML Data}}}.
	\newblock In: {\sl \bibinfo{booktitle}{Proceedings of the International
			Conference on Management of Data}}, \bibinfo{series}{SIGMOD '02},
	\bibinfo{publisher}{ACM}, pp. \bibinfo{pages}{121--132},
	\doi{10.1145/564691.564706}.
	
	\bibitemdeclare{inproceedings}{Debreceni:2014aa}
	\bibitem{Debreceni:2014aa}
	\bibinfo{author}{Csaba \surnamestart Debreceni\surnameend},
	\bibinfo{author}{\'{A}kos \surnamestart Horv\'{a}th\surnameend},
	\bibinfo{author}{\'{A}bel \surnamestart Heged\"{u}s\surnameend},
	\bibinfo{author}{Zolt\'{a}n \surnamestart Ujhelyi\surnameend},
	\bibinfo{author}{Istv\'{a}n \surnamestart R\'{a}th\surnameend} \&
	\bibinfo{author}{D\'{a}niel \surnamestart Varr\'{o}\surnameend}
	(\bibinfo{year}{2014}): \emph{\bibinfo{title}{{Query-driven Incremental
				Synchronization of View Models}}}.
	\newblock In: {\sl \bibinfo{booktitle}{Proceedings of the 2$^{nd}$ Workshop on
			View-Based, Aspect-Oriented and Orthographic Software Modelling}},
	\bibinfo{series}{VAO '14}, \bibinfo{publisher}{ACM}, pp.
	\bibinfo{pages}{31--38}, \doi{10.1145/2631675.2631677}.
	
	\bibitemdeclare{inproceedings}{Egyed:2006ic}
	\bibitem{Egyed:2006ic}
	\bibinfo{author}{Alexander \surnamestart Egyed\surnameend}
	(\bibinfo{year}{2006}): \emph{\bibinfo{title}{{Instant Consistency Checking
				for the UML}}}.
	\newblock In: {\sl \bibinfo{booktitle}{Proceedings of the 28$^{th}$
			International Conference on Software Engineering}}, \bibinfo{publisher}{ACM},
	pp. \bibinfo{pages}{381--390}, \doi{10.1145/1134285.1134339}.
	
	\bibitemdeclare{inproceedings}{Ehrig:2004aa}
	\bibitem{Ehrig:2004aa}
	\bibinfo{author}{Hartmut \surnamestart Ehrig\surnameend},
	\bibinfo{author}{Karsten \surnamestart Ehrig\surnameend},
	\bibinfo{author}{Annegret \surnamestart Habel\surnameend} \&
	\bibinfo{author}{Karl-Heinz \surnamestart Pennemann\surnameend}
	(\bibinfo{year}{2004}): \emph{\bibinfo{title}{{Constraints and Application
				Conditions: From Graphs to High-Level Structures}}}.
	\newblock In: {\sl \bibinfo{booktitle}{International Conference on Graph
			Transformations}}, {\sl \bibinfo{series}{LNCS}} \bibinfo{volume}{3256},
	\bibinfo{publisher}{Springer}, pp. \bibinfo{pages}{287--303},
	\doi{10.1007/978-3-540-30203-2\_21}.
	
	\bibitemdeclare{incollection}{Fischer:2000ut}
	\bibitem{Fischer:2000ut}
	\bibinfo{author}{Thorsten \surnamestart Fischer\surnameend},
	\bibinfo{author}{J{\"o}rg \surnamestart Niere\surnameend},
	\bibinfo{author}{Lars \surnamestart Torunski\surnameend} \&
	\bibinfo{author}{Albert \surnamestart Z{\"u}ndorf\surnameend}
	(\bibinfo{year}{2000}): \emph{\bibinfo{title}{{Story Diagrams: A New Graph
				Rewrite Language Based on the Unified Modeling Language and Java}}}.
	\newblock In: {\sl \bibinfo{booktitle}{Theory and Application of Graph
			Transformations}}, \bibinfo{publisher}{Springer}, pp.
	\bibinfo{pages}{296--309}, \doi{10.1007/978-3-540-46464-8\_21}.
	
	\bibitemdeclare{article}{Forgy:1982cg}
	\bibitem{Forgy:1982cg}
	\bibinfo{author}{Charles~L. \surnamestart Forgy\surnameend}
	(\bibinfo{year}{1982}): \emph{\bibinfo{title}{{Rete: A Fast Algorithm for the
				Many Pattern/Many object Pattern Match Problem}}}.
	\newblock {\sl \bibinfo{journal}{Artificial Intelligence}}
	\bibinfo{volume}{19}(\bibinfo{number}{1}), pp. \bibinfo{pages}{17--37},
	\doi{10.1016/0004-3702(82)90020-0}.
	
	\bibitemdeclare{book}{Gamma:1994wx}
	\bibitem{Gamma:1994wx}
	\bibinfo{author}{Erich \surnamestart Gamma\surnameend},
	\bibinfo{author}{Richard \surnamestart Helm\surnameend},
	\bibinfo{author}{Ralph \surnamestart Johnson\surnameend} \&
	\bibinfo{author}{John \surnamestart Vlissides\surnameend}
	(\bibinfo{year}{1994}): \emph{\bibinfo{title}{{Design Patterns -- Elements of
				Reusable Object-Oriented Software}}}.
	\newblock \bibinfo{publisher}{Addison-Wesley}.
	
	\bibitemdeclare{book}{Garey:1979aa}
	\bibitem{Garey:1979aa}
	\bibinfo{author}{Michael~R. \surnamestart Garey\surnameend} \&
	\bibinfo{author}{David~S. \surnamestart Johnson\surnameend}
	(\bibinfo{year}{1979}): \emph{\bibinfo{title}{{Computers and Intractability:
				A Guide to the Theory of NP-Completeness}}}.
	\newblock \bibinfo{publisher}{Freeman \& Company}.
	
	\bibitemdeclare{article}{Hanson:1996aa}
	\bibitem{Hanson:1996aa}
	\bibinfo{author}{Eric~N. \surnamestart Hanson\surnameend}
	(\bibinfo{year}{1996}): \emph{\bibinfo{title}{{The Design and Implementation
				of the Ariel Active Database Rule System}}}.
	\newblock {\sl \bibinfo{journal}{Transactions on Knowledge and Data
			Engineering}} \bibinfo{volume}{8}(\bibinfo{number}{1}), pp.
	\bibinfo{pages}{157--172}, \doi{10.1109/69.485644}.
	
	\bibitemdeclare{article}{Hanson:2002aa}
	\bibitem{Hanson:2002aa}
	\bibinfo{author}{Eric~N. \surnamestart Hanson\surnameend},
	\bibinfo{author}{Sreenath \surnamestart Bodagala\surnameend} \&
	\bibinfo{author}{Ullas \surnamestart Chadaga\surnameend}
	(\bibinfo{year}{2002}): \emph{\bibinfo{title}{{Trigger Condition Testing and
				View Maintenance Using Optimized Discrimination Networks}}}.
	\newblock {\sl \bibinfo{journal}{Transactions on Knowledge and Data
			Engineering}} \bibinfo{volume}{14}(\bibinfo{number}{2}), pp.
	\bibinfo{pages}{261--280}, \doi{10.1109/69.991716}.
	
	\bibitemdeclare{inproceedings}{Harrison:1992aa}
	\bibitem{Harrison:1992aa}
	\bibinfo{author}{John~V. \surnamestart Harrison\surnameend} \&
	\bibinfo{author}{Suzanne~W. \surnamestart Dietrich\surnameend}
	(\bibinfo{year}{1992}): \emph{\bibinfo{title}{{Maintenance of Materialized
				Views in a Deductive Database: An Update Propagation Approach}}}.
	\newblock In: {\sl \bibinfo{booktitle}{Workshop on Deductive Databases}},
	\bibinfo{publisher}{JICSLP}, pp. \bibinfo{pages}{56--65}.
	
	\bibitemdeclare{inproceedings}{Kiesel:1993aa}
	\bibitem{Kiesel:1993aa}
	\bibinfo{author}{Norbert \surnamestart Kiesel\surnameend},
	\bibinfo{author}{Andy \surnamestart Sch\"{u}rr\surnameend} \&
	\bibinfo{author}{Bernhard \surnamestart Westfechtel\surnameend}
	(\bibinfo{year}{1993}): \emph{\bibinfo{title}{{GRAS, a graph-oriented
				database system for (software) engineering applications}}}.
	\newblock In: {\sl \bibinfo{booktitle}{Proceeding of the 6$^{th}$ International
			Workshop on Computer-Aided Software Engineering}}, \bibinfo{publisher}{IEEE},
	pp. \bibinfo{pages}{272--286}, \doi{10.1109/CASE.1993.634829}.
	
	\bibitemdeclare{article}{Lee:1992rt}
	\bibitem{Lee:1992rt}
	\bibinfo{author}{Ho~Soo \surnamestart Lee\surnameend} \&
	\bibinfo{author}{Marshall~I. \surnamestart Schor\surnameend}
	(\bibinfo{year}{1992}): \emph{\bibinfo{title}{{Match Algorithms for
				Generalized Rete Networks}}}.
	\newblock {\sl \bibinfo{journal}{Artificial Intelligence}}
	\bibinfo{volume}{54}(\bibinfo{number}{2}), pp. \bibinfo{pages}{249--274},
	\doi{10.1016/0004-3702(92)90047-2}.
	
	\bibitemdeclare{inproceedings}{Liu:2000aa}
	\bibitem{Liu:2000aa}
	\bibinfo{author}{Jixue \surnamestart Liu\surnameend}, \bibinfo{author}{Millist
		\surnamestart Vincent\surnameend} \& \bibinfo{author}{Mukesh \surnamestart
		Mohania\surnameend} (\bibinfo{year}{2000}):
	\emph{\bibinfo{title}{{Maintaining Views in Object-Relational Databases}}}.
	\newblock In: {\sl \bibinfo{booktitle}{Proceedings of the $9^{th}$
			International Conference on Information and Knowledge Management}},
	\bibinfo{series}{CIKM '00}, \bibinfo{publisher}{ACM}, pp.
	\bibinfo{pages}{102--109}, \doi{10.1145/354756.354807}.
	
	\bibitemdeclare{article}{Lucredio:2010ew}
	\bibitem{Lucredio:2010ew}
	\bibinfo{author}{Daniel \surnamestart Lucr{\'e}dio\surnameend},
	\bibinfo{author}{Renata \surnamestart Fortes\surnameend} \&
	\bibinfo{author}{Jon \surnamestart Whittle\surnameend}
	(\bibinfo{year}{2010}): \emph{\bibinfo{title}{{MOOGLE: a metamodel-based
				model search engine}}}.
	\newblock {\sl \bibinfo{journal}{Software {\&} Systems Modeling}}
	\bibinfo{volume}{11}(\bibinfo{number}{2}), pp. \bibinfo{pages}{183--208},
	\doi{10.1007/s10270-010-0167-7}.
	
	\bibitemdeclare{inproceedings}{Miranker:1987tr}
	\bibitem{Miranker:1987tr}
	\bibinfo{author}{Daniel~P. \surnamestart Miranker\surnameend}
	(\bibinfo{year}{1987}): \emph{\bibinfo{title}{{TREAT: A Better Match
				Algorithm for AI Production Systems}}}.
	\newblock In: {\sl \bibinfo{booktitle}{{Proceedings of the 6$^{th}$ National
				Conference on Artificial Intelligence}}}, \bibinfo{volume}{1},
	\bibinfo{publisher}{AAAI Press}, pp. \bibinfo{pages}{42--47}.
	
	\bibitemdeclare{inproceedings}{Niere:2003uq}
	\bibitem{Niere:2003uq}
	\bibinfo{author}{J{\"o}rg \surnamestart Niere\surnameend},
	\bibinfo{author}{J{\"o}rg \surnamestart Wadsack\surnameend} \&
	\bibinfo{author}{Lothar \surnamestart Wendehals\surnameend}
	(\bibinfo{year}{2003}): \emph{\bibinfo{title}{{Handling large search space in
				pattern-based reverse engineering}}}.
	\newblock In: {\sl \bibinfo{booktitle}{Proceedings of the 11$^{th}$
			International Workshop on Program Comprehension}}, \bibinfo{publisher}{IEEE},
	pp. \bibinfo{pages}{274--279}, \doi{10.1109/WPC.2003.1199212}.
	
	\bibitemdeclare{article}{Qian:1991aa}
	\bibitem{Qian:1991aa}
	\bibinfo{author}{Xiaolei \surnamestart Qian\surnameend} \& \bibinfo{author}{Gio
		\surnamestart Wiederhold\surnameend} (\bibinfo{year}{1991}):
	\emph{\bibinfo{title}{{Incremental Recomputation of Active Relational
				Expressions}}}.
	\newblock {\sl \bibinfo{journal}{Transactions on Knowledge and Data
			Engineering}} \bibinfo{volume}{3}(\bibinfo{number}{3}), pp.
	\bibinfo{pages}{337--341}, \doi{10.1109/69.91063}.
	
	\bibitemdeclare{inproceedings}{Rath:2008ex}
	\bibitem{Rath:2008ex}
	\bibinfo{author}{Istv{\'a}n \surnamestart R{\'a}th\surnameend},
	\bibinfo{author}{G{\'a}bor \surnamestart Bergmann\surnameend},
	\bibinfo{author}{Andr{\'a}s \surnamestart {\"O}kr{\"o}s\surnameend} \&
	\bibinfo{author}{D{\'a}niel \surnamestart Varr{\'o}\surnameend}
	(\bibinfo{year}{2008}): \emph{\bibinfo{title}{{Live Model Transformations
				Driven by Incremental Pattern Matching}}}.
	\newblock In: {\sl \bibinfo{booktitle}{Proceedings of the 6$^{th}$
			International Conference on Theory and Practice of Model Transformations}},
	\bibinfo{publisher}{Springer}, pp. \bibinfo{pages}{107--121},
	\doi{10.1007/978-3-540-69927-9\_8}.
	
	\bibitemdeclare{inproceedings}{Rath:2012aa}
	\bibitem{Rath:2012aa}
	\bibinfo{author}{Istv\'{a}n \surnamestart R\'{a}th\surnameend},
	\bibinfo{author}{\'{A}bel \surnamestart Heged\"{u}s\surnameend} \&
	\bibinfo{author}{D\'{a}niel \surnamestart Varr\'{o}\surnameend}
	(\bibinfo{year}{2012}): \emph{\bibinfo{title}{{Derived Features for EMF by
				Integrating Advanced Model Queries}}}.
	\newblock In: {\sl \bibinfo{booktitle}{Proceedings of the 8$^{th}$ European
			Conference on Modelling Foundations and Applications}},
	\bibinfo{series}{ECMFA'12}, \bibinfo{publisher}{Springer}, pp.
	\bibinfo{pages}{102--117}, \doi{10.1007/978-3-642-31491-9\_10}.
	
	\bibitemdeclare{book}{Robinson:2015aa}
	\bibitem{Robinson:2015aa}
	\bibinfo{author}{Ian \surnamestart Robinson\surnameend}, \bibinfo{author}{Jim
		\surnamestart Webber\surnameend} \& \bibinfo{author}{Emil \surnamestart
		Eifrem\surnameend} (\bibinfo{year}{2015}): \emph{\bibinfo{title}{{Graph
				Databases (Second Edition)}}}.
	\newblock \bibinfo{publisher}{O'Reilly Media}.
	
	\bibitemdeclare{article}{Rodriguez:2010aa}
	\bibitem{Rodriguez:2010aa}
	\bibinfo{author}{Marko~A. \surnamestart Rodriguez\surnameend} \&
	\bibinfo{author}{Peter \surnamestart Neubauer\surnameend}
	(\bibinfo{year}{2010}): \emph{\bibinfo{title}{{The Graph Traversal
				Pattern}}}.
	\newblock {\sl \bibinfo{journal}{CoRR Journal}}
	\bibinfo{volume}{1004}(\bibinfo{number}{1001}).
	
	\bibitemdeclare{article}{Seibel:2010jb}
	\bibitem{Seibel:2010jb}
	\bibinfo{author}{Andreas \surnamestart Seibel\surnameend},
	\bibinfo{author}{Stefan \surnamestart Neumann\surnameend} \&
	\bibinfo{author}{Holger \surnamestart Giese\surnameend}
	(\bibinfo{year}{2010}): \emph{\bibinfo{title}{{Dynamic hierarchical mega
				models: comprehensive traceability and its efficient maintenance}}}.
	\newblock {\sl \bibinfo{journal}{Software {\&} Systems Modeling}}
	\bibinfo{volume}{9}(\bibinfo{number}{4}), pp. \bibinfo{pages}{493--528},
	\doi{10.1007/s10270-009-0146-z}.
	
	\bibitemdeclare{inproceedings}{Srinivasa:2005ab}
	\bibitem{Srinivasa:2005ab}
	\bibinfo{author}{Srinath \surnamestart Srinivasa\surnameend} \&
	\bibinfo{author}{Martin \surnamestart Maier\surnameend}
	(\bibinfo{year}{2005}): \emph{\bibinfo{title}{{LWI and Safari: A New Index
				Structure and Query Model for Graph Databases}}}.
	\newblock In: {\sl \bibinfo{booktitle}{Proceedings of the 11$^{th}$
			International Conference on Management of Data}},
	\bibinfo{publisher}{Computer Society of India}.
	
	\bibitemdeclare{inproceedings}{Varro:2013bu}
	\bibitem{Varro:2013bu}
	\bibinfo{author}{Gergely \surnamestart Varr\'{o}\surnameend} \&
	\bibinfo{author}{Frederik \surnamestart Deckwerth\surnameend}
	(\bibinfo{year}{2013}): \emph{\bibinfo{title}{{A Rete Network Construction
				Algorithm for Incremental Pattern Matching}}}.
	\newblock In: {\sl \bibinfo{booktitle}{Theory and Practice of Model
			Transformations}}, {\sl \bibinfo{series}{LNCS}} \bibinfo{volume}{7909},
	\bibinfo{publisher}{Springer}, pp. \bibinfo{pages}{125--140},
	\doi{10.1007/978-3-642-38883-5\_13}.
	
	\bibitemdeclare{article}{Yan:2005aa}
	\bibitem{Yan:2005aa}
	\bibinfo{author}{Xifeng \surnamestart Yan\surnameend},
	\bibinfo{author}{Philip~S. \surnamestart Yu\surnameend} \&
	\bibinfo{author}{Jiawei \surnamestart Han\surnameend} (\bibinfo{year}{2005}):
	\emph{\bibinfo{title}{{Graph Indexing Based on Discriminative Frequent
				Structure Analysis}}}.
	\newblock {\sl \bibinfo{journal}{Transactions on Database Systems}}
	\bibinfo{volume}{30}(\bibinfo{number}{4}), pp. \bibinfo{pages}{960--993},
	\doi{10.1145/1114244.1114248}.
	
	\bibitemdeclare{inproceedings}{Zhuge:1998aa}
	\bibitem{Zhuge:1998aa}
	\bibinfo{author}{Yue \surnamestart Zhuge\surnameend} \&
	\bibinfo{author}{H.~\surnamestart Garcia-Molina\surnameend}
	(\bibinfo{year}{1998}): \emph{\bibinfo{title}{{Graph Structured Views and
				Their Incremental Maintenance}}}.
	\newblock In: {\sl \bibinfo{booktitle}{Proceedings of the 14$^{th}$
			International Conference on Data Engineering}}, \bibinfo{publisher}{IEEE},
	pp. \bibinfo{pages}{116--125}, \doi{10.1109/ICDE.1998.655767}.
\end{thebibliography}

\end{document}